\documentclass[3p]{article}
\usepackage[utf8]{inputenc}

\usepackage{relsize}
\usepackage{appendix}
\usepackage{amsmath}
\usepackage{graphicx}
\usepackage{lineno}
\usepackage{array}
\usepackage{longtable}
\usepackage[utf8]{inputenc}
\usepackage{physics}
\usepackage{subcaption}
\usepackage{graphicx} 
\usepackage{siunitx}
\usepackage{comment}
\usepackage{empheq}
\usepackage[shortlabels]{enumitem} 
\usepackage[notextcomp]{stix} 

\usepackage{smartdiagram}
\usepackage{pgfplots}
\usepackage[hidelinks]{hyperref}

\usepackage{standalone}

\def\isplot{1} 

\usepackage{subcaption}
\usepackage{pgf,tikz,pgfplots}
 \usetikzlibrary{math} 
   \usepackage{pgffor} 
\usepackage{amsmath,amssymb,braket,}
\usepackage{pst-plot} 
\usetikzlibrary{spy,arrows}
\usetikzlibrary{shapes,arrows,positioning,backgrounds,fit,calc,math,tikzmark}
\usetikzlibrary{decorations.pathmorphing}\usetikzlibrary{decorations.pathreplacing}\usetikzlibrary{decorations.shapes}\usetikzlibrary{decorations.text}\usetikzlibrary{decorations.markings}\usetikzlibrary{decorations.fractals}\usetikzlibrary{decorations.footprints}
\tikzset{cross/.style={cross out, draw=black, minimum size=2*(#1-\pgflinewidth), inner sep=0pt, outer sep=0pt},
cross/.default={3pt}}
\usetikzlibrary{decorations}
\usetikzlibrary{plotmarks}
\usetikzlibrary{patterns}

  \pgfplotsset{every tick label/.append style={font=\tiny}}

  \pgfplotsset{every x tick label/.append style={font=\tiny, yshift=0.5ex}}

  \pgfplotsset{every y tick label/.append style={font=\tiny, xshift=0.5ex}}

\usepackage{ifthen}

  \usetikzlibrary{math} 
   \usepackage{pgffor} 

\tikzset{cross/.style={cross out, draw=black, minimum size=2*(#1-\pgflinewidth), inner sep=0pt, outer sep=0pt},
cross/.default={3pt}}

\tikzset{->-/.style={decoration={
  markings,
  mark=at position #1 with {\arrow{>}}},postaction={decorate}}}

\definecolor{greenmf}{RGB}{0,128,0}
\definecolor{redmf}{RGB}{128,0,0}

\definecolor{blueedf}{RGB}{0,91,187}
\definecolor{bluededf}{RGB}{9,53,122}
\definecolor{blueledf}{RGB}{0,206,209}

\definecolor{orangededf}{RGB}{250,88,21}
\definecolor{orangeedf}{RGB}{240,160,47}
\definecolor{orangeledf}{RGB}{0,206,209}

\definecolor{greenedf}{RGB}{80,158,47}
\definecolor{greendedf}{RGB}{0,100,0}
\definecolor{greenledf}{RGB}{196,214,47}

\definecolor{indigo}{RGB}{106,44,145}
\definecolor{gold}{RGB}{240,215,0}

\definecolor{myanalcolor}{RGB}{0,0,0}
\definecolor{myrelcolor}{RGB}{180,180,180}
\definecolor{myrelcolor2}{RGB}{240,215,0}

\definecolor{mycolor1a}{RGB}{80,158,47}
\definecolor{mycolor1b}{RGB}{0,100,0} 
\definecolor{mycolor1c}{RGB}{196,214,0} 

\definecolor{mycolor2a}{RGB}{0,91,187}
\definecolor{mycolor2b}{RGB}{9,53,122}
\definecolor{mycolor2c}{RGB}{0,206,209}

\definecolor{mycolor3a}{RGB}{250,0,0}
\definecolor{mycolor3b}{RGB}{148,0,211}
\definecolor{mycolor3c}{RGB}{240,160,47}

\newcommand{\vect}[1]{\underline{#1}}
\newcommand{\mtens}[1]{\underline{\underline{#1}}}
\newcommand{\mgrad}{\vect{\grad}}
\newcommand{\mdiv}{\mathrm{div}}

\renewcommand{\lim}[2]{\underset{#1 \xrightarrow[]{} #2}{\mathrm{lim}}}

\newcommand{\lra}[1]{{\langle #1 \rangle}}

\newcommand{\eref}[1]{Eq.~\eqref{eq:#1}}
\newcommand{\iref}[1]{\ref{item:#1}}
\newcommand{\tref}[1]{\figurename{}~\ref{tikz:#1}}
\newcommand{\fref}[1]{\figurename{}~\ref{fig:#1}}
\newcommand{\sref}[1]{Sect.~\ref{sec:#1}}
\newcommand{\aref}[1]{Appendix~\ref{app:#1}}

\makeatletter
\newcommand{\labitem}[2]{%
\def\@itemlabel{#1}
\item
\def\@currentlabel{#1}\label{#2}}
\makeatother

\begin{document}

\title{
Analysis of wall-modelled particle/mesh PDF methods for turbulent parietal flows
}%
\date{}

\maketitle
\vspace{1cm}
\begin{center}
    
\author{\large\textbf{G. Balvet$^{*,1,2}$, J.-P. Minier$^{1,2}$, Y. Roustan$^{2}$, M. Ferrand$^{1,2}$}
}~\\
\vspace{1cm}

$^*$ contact address: guilhem.balvet@edf.fr~\\
$^{1}$ EDF R\&D, Fluid Mechanics, Energy and Environment Dept., 6 Quai Watier,78400, Chatou, France~\\
$^{2}$ CEREA, \'Ecole des Ponts, Île-de-France, France  ~\\
\end{center}

DOI: https://doi.org/10.1515/mcma-2023-2017

\vspace{1cm}

\begin{abstract}
{
 Lagrangian stochastic methods are widely used to model turbulent flows. Scarce consideration has, however, been devoted to the treatment of the near-wall region and to the formulation of a proper wall-boundary condition. With respect to this issue, the main purpose of this paper is to present an in-depth analysis of such flows when relying on particle/mesh formulations of the probability density function (PDF) model.
 This is translated into three objectives. The first objective is to assess the existing anelastic wall-boundary condition and present new validation results. The second objective is to analyze the impact of the interpolation of the mean fields at particle positions on their dynamics. The third objective is to investigate the spatial error affecting covariance estimators when they are extracted on coarse volumes. All these developments allow to ascertain that the key dynamical statistics of wall-bounded flows are properly captured even for coarse spatial resolutions. ~\\

 MSC 2010 class: 76M35 ~\\

\textbf{keyword: Interpolation scheme error, Lagrangian stochastic methods, Particle/mesh methods, Statistic estimation, Wall-boundary condition }

}
\end{abstract}
\maketitle

\vspace{1cm}



\section{Introduction}~

In Lagrangian stochastic methods, also referred to as Lagrangian probability density function (PDF) methods, the PDF associated to selected variables of interest, which form the particle state vector, is estimated, in a weak sense, by Monte Carlo methods from a large number of `stochastic particles'~\cite{pope_1985}. These approaches belong to the category of reduced statistical descriptions~\cite{minier2016statistical} in which a Lagrangian standpoint is adopted to model and simulate single-phase, as well as poly-disperse two-phase, turbulent flows~\cite{pope2000turbulent,minier2001PDF} and where the eliminated degrees of freedom are replaced by stochastic models. As indicated by their name, one-particle PDF models have an intrinsic Lagrangian nature from which one-point one-time Eulerian PDFs are automatically derived, using a one-way-street procedure, to obtain mean fields. The theoretical framework supporting the approach is now well-established for single-phase turbulent flows~\cite{pope_1985,pope2000turbulent}, as well as for disperse two-phase ones~\cite{minier2001PDF,minier2015lagrangian}, which means that we can safely concentrate our attention on the physical and numerical issues related to the precise formulation of such particle-based stochastic models.

The ability of PDF methods to treat without approximation convection as well as chemical source terms (and, more generally, the mean value of any function, however complex or non-linear, of the variables entering the particle state vector) makes them attractive candidates to simulate pollutant dispersion and/or reactive flows \cite{sabelfeld2012random}. This has long been recognized in the atmospheric~\cite{rodean1996,wilson_review_1996, Thomson_Wilson_2013} and combustion~\cite{pope_relationship_1979,pope1981monte,pope_lagrangian_1983} communities. In the course of the development of Lagrangian stochastic models, two concerns have, however, often surfaced. The first concern is related to the respect to the well-mixed condition which states that, for an incompressible flow, a uniform distribution of particle should remain so~\cite{sawford_generalized_1986,thomson_criteria_1987}. Indeed, since fluid particles represent the same fixed amount of mass, preserving a uniform concentration is equivalent to stating that the mass conservation equation should be satisfied. Actually, this issue was clarified very early~\cite{pope_1985,pope_consistency_1987,mcinnes1992phys} and has been repeated in several works~\cite{minier2001PDF,minier2014guidelines}, where it was demonstrated that, as long as the mean pressure gradient is properly introduced in the particle velocity equation, the well-mixed criterion is automatically satisfied. This was revisited recently in the analysis of \cite{bahlali_2020_hyb_meth} with a view towards atmospheric applications which confirmed previous conclusions. The second concern is related to the wall-boundary condition we should apply to ensure that the `law of the wall' \cite{pope2000turbulent} is correctly reproduced by Lagrangian simulations. Contrary to the first concern, this point has received less attention (see an overview of existing attempts in~\cite{haworth_progress_2010}) and is still the subject of some confusion as to the form and the physical meaning of the wall-boundary condition for fluid particles. In many applications, a simple elastic condition is applied at the wall boundary. Such a condition is clearly wrong for particle streamwise velocity components as it cancels the exchange of momentum occurring in the near-wall region, which is a key characteristic of the physics of wall boundary layers~\cite{pope2000turbulent}. On the other hand, an an-elastic wall boundary conditions was proposed in earlier works~\cite{Dreeben1997ProbabilityDF,minier1999wall}, though its significance may not have yet been perceived for stochastic-particle-based simulations. The issue of what wall-boundary condition should be enforced in turbulent wall-bounded flows was addressed in~\cite{bahlali_2020_hyb_meth} which revealed that an elastic rebound condition leads to serious errors in the near-wall region and brought further validation for the an-elastic boundary condition. Yet, the analysis remained incomplete as to whether the correct profiles of the logarithmic region were really retrieved but was helpful to bring out a number of numerical issues concerning the interpolation of mean fields at particle positions and how particle statistics are to be calculated. In that sense, the present work is a follow-up of this first study and aims at clarifying the issue of the wall-boundary conditions needed in the spirit of the wall-function treatment of turbulent boundary layers, as well as bringing insights into the physical and numerical issues involved.

To address the above-mentioned issues, we analyse numerical outcomes in a neutral surface boundary layer. Since this configuration is the most classical way to describe near-ground atmospheric flows in the absence of thermal effects, it is of first importance for atmospheric applications. This is also a situation where the wall-boundary condition plays a key role. Furthermore, analytical results are available which allows to monitor numerical errors. Finally, this situation remains simple enough to allow in-depth numerical investigations to be performed, while conclusions remain applicable in more complex geometries since it is applied to describe locally a turbulent flow in the immediate vicinity of small wall-surface elements. In contrast to the first studies on the an-elastic wall-boundary condition~\cite{Dreeben1997ProbabilityDF,minier1999wall} which were carried out using stand-alone simulations, present results were obtained using a hybrid finite-volume/Particle (FV/Particle) numerical method, corresponding to a hybrid Moments/PDF description of turbulent flows, and are therefore interesting to assess since they complement these first studies and provide additional support.

In short, the present work has a three-fold objective:
\begin{enumerate}[(i)]
\item to present new numerical results to validate the an-elastic boundary condition and point out the shortcomings of the often applied elastic condition;
\item to investigate the numerical errors induced when interpolating mean fields at particle locations and propose local schemes to simulate particle dynamics;
\item to bring out statistical artefacts when extracting particle statistics in volumes where the local homogeneity assumption fails and to propose correction terms.
\end{enumerate}

This paper is organised as follows. In~\sref{background}~, we revisit the main aspects of Lagrangian stochastic methods, the an-elastic wall-boundary conditions, as well as the numerical hybrid formulation used in later sections. Then, detailed numerical results are presented in~\sref{verif_bc} to validate the an-elastic wall-boundary condition. The issues related to the interpolation of mean fields at particle locations are addressed in~\sref{interpolation_issue}, while a careful investigation of potential artefacts in the statistical treatment of particle dynamics is carried out in~\sref{sp_error_stat}. Conclusions are then given in~\sref{conclusions}.

\newpage
\section{Background on the Lagrangian Stochastic Methods }\label{sec:background}

\subsection{Lagrangian PDF Modelling of Turbulent Flows}~

This section presents a brief background on Lagrangian stochastic methods used to model turbulent flows. There is now a rather wide literature dedicated to presenting the PDF approach both for single-phase~\cite{pope_1985,pope2000turbulent,haworth_progress_2010} as well as for disperse two-phase flows~\cite{minier2001PDF,minier2015lagrangian,minier2016statistical}, so that only key points are recalled below concentrating on single-phase turbulent flows.

As indicated by their name, these methods correspond to a statistical description which aims at modelling the PDF of a number of selected variables of interest for a given turbulent flow \cite{sabelfeld2012random}. Using a Lagrangian formulation to model and simulate such flows means that we are tracking the evolution of a large number of numerical or notional particles in the computational domain. The instantaneous variables attached to these particles make up the particle state vector and each particle can then be seen as an independent realization, or sample, of the corresponding PDF. In the present study, we consider single-phase incompressible high-Reynolds-number flows and the retained particle state vector $\vect{Z}(t)$ is made up by the particle instantaneous position $( \vect{X}(t))$ and velocity $( \vect{U}(t))$, i.e. $\vect{Z}(t) = ( \vect{X}(t), \vect{U}(t))$ where $t$ is the time.

Once the state vector is defined, its time evolution has to be modelled. For reasons discussed at length in several works~\cite{pope_1985,pope2000turbulent,minier2001PDF,minier2016statistical}, there are strong physical arguments, based on the Kolmogorov theory, to suggest to model the time evolution of the particle state vector by a general stochastic diffusion. This means that the increments of $\vect{Z}(t)$ over a small time increments $\dd t$ have the form~\cite{gardiner_handbook_1985,ottinger_stochastic_1996}:
\begin{equation}\label{eq:dif_process}
       \dd \vect{Z} = \vect{\mathcal{A}}\left(t;\vect{Z}(t),\lra{\mathcal{F}(\vect{Z}(t))}\right) \dd t +  \mtens{\mathcal{B}}\, \left(t;\vect{Z}(t),\lra{\mathcal{G}(\vect{Z}(t))}\right) \dd \vect{W}.
\end{equation}
The right-hand-side (RHS of Eq.~\eqref{eq:dif_process} involves two terms having different physical meanings. On the one hand, the deterministic term in factor of the time increments is specified through the drift vector $\vect {\mathcal{A}}$ which governs the linear-in-$\dd t$ evolution of the conditional mean increments of $\dd \vect{Z}$~\cite{minier2015lagrangian}. On the other hand, the stochastic term in front of the Wiener process increments $ \dd \vect{W}$ is specified through the diffusion matrix $\mtens{\mathcal{B}}$ which characterizes the linear-in-$\dd t$ evolution of the conditional variances around these mean increments~\cite{minier2015lagrangian}. In the drift vector and the diffusion matrix, a possible dependence on mean fields extracted from the PDF, manifested by the terms $\lra{\mathcal{F}(\vect{Z}(t))}$ and $\lra{\mathcal{G}(\vect{Z}(t))}$, is included, where $\lra{(.)}$ stands for the averaging operator. The Wiener process $W$ is a stochastic process with continuous, yet nowhere differentiable, trajectories and whose increments $\dd W$ over a small time increment $\dd t$ satisfy $\lra{\dd W}=0$ and $\lra{(\dd W)^2}=\dd t$, so that it accounts for white-noise effects. The diffusion term in Eq.~\eqref{eq:dif_process} is written as a function of $\dd \vect{W}$ which is a vector of independent Wiener processes. In the corresponding sample space, the PDF follows a Fokker-Planck equation~\cite{pope_1985,pope2000turbulent,ottinger_stochastic_1996} and solving the stochastic differential equations (SDEs) in Eq.~\eqref{eq:dif_process} for a large number of particles is equivalent, in a weak sense, to solving the corresponding Fokker-Planck equation through Monte Carlo methods (see detailed presentations in~\cite[Chapter 1]{chibbaro2014stochastic}). In practice, this means that we are concerned with statistics derived from a set of particles rather than with the history of a single one.

In the present work, the evolution of the retained particle state vector $\vect{Z}(t) = ( \vect{X}(t), \vect{U}(t))$ is modelled with the simplified Langevin model (SLM) which, as indicated by its name, is the simplest formulation in the class of generalized Langevin models (GLMs) developed over the years by Pope and co-workers~\cite{pope_1985,haworth1986generalized,pope2000turbulent}. The SLM writes as
\begin{subequations}
\begin{align}
 \dd \vect{X} & = \vect{U} \dd t , \label{rq:SLM_cont_X}\\
 \dd \vect{U} & = - \dfrac{1}{\rho }\mgrad \lra{P}\left(t;\vect{X}(t)\right)\dd t - \dfrac{\vect{U} - \lra{\vect{U}(t;\vect{X}(t))}}{C_L \frac{k(t;\vect{X}(t))}{\epsilon(t;\vect{X}(t))}}\dd t+ \sqrt{ C_0 \epsilon(t;\vect{X}(t)) } \, \dd \vect{W}~. \label{eq:SLM_cont_U}
\end{align}
\label{eq:SLM_cont}
\end{subequations}

In \eref{SLM_cont_U}, the first term corresponds to the effect of the mean pressure gradient, with $\lra{P}$ the mean pressure and $\rho$ the density of the fluid. The second term is a return-to-equilibrium term written as a relaxation toward the local value of the mean velocity field $\lra{\vect{U}}$ at the particle location. Note that the presence of the mean velocity field in the drift term is an example of the dependence on the PDF of $\vect{Z}(t)$ indicated as $\lra{\mathcal{F}(\vect{Z}(t))}$ in Eq.~\eqref{eq:dif_process}. The relaxation term involves the Lagrangian time scale, which leaving out the dependence on $(t;\vect{X}(t))$ for the sake of keeping simple notations, is:
\begin{equation}
T_L= C_L \frac{k}{\epsilon},
\label{eq:def_TL}
\end{equation}
where $k$ is the turbulence kinetic energy of the fluid ($k=1/2\,\lra{(\vect{u}\cdot \vect{u})}$, where an instantaneous variable written with an upper-case letter can be decomposed into a mean component noted between bracket and a fluctuating one written with a lower-case letter, such as $\vect{U} = \lra{\vect{U}} + \vect{u}$) and $\epsilon$ its dissipation rate. The diffusion term in Eq.~\eqref{eq:SLM_cont_U} involves the Kolmogorov constant $C_0$ whose value is taken here as equal to 3.5 \cite{minier1999wall}. In order to retrieve the correct equation for the kinetic energy budget, the constant $C_L$ in Eq.~\eqref{eq:def_TL} is given by $C_L = (\tfrac{1}{2} + \tfrac{3}{4} C_0)^{-1}$.
Finally, since molecular viscous effects are not accounted in Eq.~\eqref{eq:SLM_cont}, this form of the SLM model is valid only in the high-Reynolds-number limit. To simulate wall-bounded flows, it then necessary to develop specific wall-boundary conditions, as is done in~\sref{wall_bc}.

\subsection{Reference An-elastic Wall-boundary condition}\label{sec:wall_bc}~

For high-Reynolds-number parietal flows, the development of wall-boundary conditions in the spirit of the wall-function approach is important. Indeed, even though some Lagrangian stochastic models can simulate low-Reynolds-number flows (see e.g.~\cite{Dreeben1997ProbabilityDF, dreeben_pope_1998,Waclawczyk_minier_2004_near_wall}), the highly inhomogeneous and anisotropic variations occurring in the viscous sub-layer require a costly refinement in this area and remain a challenging and time-consuming issue. Therefore, similarly to the wall-function treatment used in moment approaches, the wall-boundary condition for particles aims to reproduce the behaviour of the logarithmic zone without considering explicitly the presence of this viscous sub-layer.

In classical FV formulations, boundary conditions are applied for the different mean fields at the centre of the first cell. The latter one must then be located within the logarithmic zone. In particle methods, the information is made up by instantaneous variables (e.g. instantaneous velocities) carried along particle trajectories whose location can be arbitrary close to the surface. Thus, the boundary surface considered for particles should also be in the logarithmic zone. To fulfil this condition, the boundary condition is then applied at a plane locally parallel to the physical wall but slightly shifted to a distance $z_{pl}$, as shown in~\tref{rebound}. The issue is to express what condition should be imposed on instantaneous particle-attached variables to obtain the correct resulting statistics which are representative of the physics of the logarithmic zone.
\begin{figure}[h!]
    \centering
    \includegraphics[width = 0.5 \linewidth]{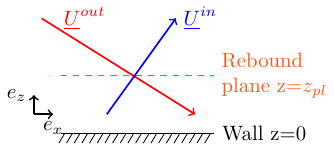}
    \caption{Illustration of the Lagrangian wall-boundary condition. The boundary condition is applied at a rebound plane shifted from the wall by a distance $z_{pl}$. For each particle crossing the rebound plane and leaving the domain a particle is reflected with properties estimated to respect statistical condition at the rebound plane (zero mean normal velocity, proper shear stress, ...).}
    \label{tikz:rebound}
\end{figure}

Only particles inside the domain, i.e. above the rebound plane on the scheme~\tref{rebound}, are simulated. Thus, for each particle leaving the domain with instantaneous properties $\vect{Z}^{out}$, a reflected mirror particle is injected with the instantaneous properties $\vect{Z}^{in}$ as represented in~\tref{rebound}. In order to impose conditions on the statistics at the rebound plane $\langle f(\vect{Z})\rangle^{pl}$ (for example non permeability of the wall, shear stress, scalar fluxes, ...), the instantaneous properties of the particles injected are chosen based on the corresponding values for the outgoing particles by considering requirements for the resulting fluxes at the rebound plane:

\begin{equation}
    \langle f(\vect{Z})\rangle^{pl}  =  \frac{1}{2} \left(\langle f(\vect{Z}) \rangle^{in} +\langle f(\vect{Z}) \rangle^{out} \right).
    \label{eq:bc_averaged}
\end{equation}

First, in the logarithmic zone near the wall, the mean velocity normal to the wall is null: $\lra{W}^{pl} = 0$. This condition is necessary to ensure the non-permeability of the wall, thus avoiding spurious accumulation or depletion of particles in its vicinity. This can easily be respected by imposing this condition on each pair of leaving/entering particles. The normal velocity of the injected particle $ W^{in}$ is then the opposite of the normal velocity of the corresponding outgoing particle $W^{out}$:
\begin{equation}\label{eq:bc_w}
    W^{in} = - W^{out}~,
\end{equation}
from which it follows that specular reflection is applied to the particle distance to the wall (here the vertical coordinate $z$):
\begin{equation}\label{eq:bc_z}
    z^{in} = 2\,z_{pl}- z^{out}~.
\end{equation}

The condition on the streamwise velocity is less straightforward. Too often, a specular rebound is imposed with no specific condition for this variable. If this choice respects the necessity to keep the particles within the domain, it does not respect the physics considered. Indeed this amounts to imposing a zero gradient condition in the streamwise direction. This is clearly at variance with having a constant shear stress and with the physical exchange of momentum characteristic of the logarithmic zone. With the purpose of conserving this exchange of momentum and the gaussianity of the stochastic increments,~\cite{dreeben1997wall} and~\cite{minier1999wall} proposed to estimate the injected streamwise velocity $U^{in}$ as a linear function of $U^{out}$ and $W^{out}$:

\begin{equation}\label{eq:model_uin}
     U^{in} = U^{out} + \alpha W^{out} .
\end{equation}

The value $\alpha$ is then determined so that the correct shear stress at the plane is retrieved. This yields:

\begin{equation}\label{eq:val_alpha}
    \alpha = -2  \frac{\langle uw \rangle (z_{pl})}{ \langle w^2 \rangle(z_{pl})}.
\end{equation}

It is worth noting that a similar formulation is correct in the three directions replacing $\lra{uw}$ by $\mtens{R}\cdot \vect{n}$ where $\mtens{R}$ is the Reynolds tensor with $R_{ij} = \lra{u_iu_j}$ and $\vect{n}$ is the vector normal to the face. Thus the general form of the boundary condition imposed on the velocity is given in~\eref{good_bc}:

 \begin{equation}\label{eq:good_bc}
    \vect{U}^{in} = \vect{U}^{out} - 2 \frac{ \mtens{R}\cdot \vect{n}}{ R_{nn} } \vect{U}^{out} \cdot \vect{n}.
\end{equation}

This an-elastic rebound condition is considered as the reference one. It will be compared to the elastic or specular rebound and validated on both smooth and coarse walls in~\sref{verif_bc_sm_rg_wall}.

\subsection{Numerical Formulation}\label{sec:numer_methods}~

In practice, Lagrangian PDF models are implemented as particle/mesh methods. In these methods, the algorithm followed can be decomposed into four numerical steps: the estimation of the mean carrier fields on a mesh; the interpolation of these mean carrier fields from the mesh to the particle locations; the temporal integration of the instantaneous quantities associated to the particles; the estimation of the statistics extracted from the set of particles. In this section, these steps and the numerical choice considered are briefly presented.  A sketch of the overall formulation of the present hybrid method is shown in~\tref{num_meth_sum} and is used to introduce the key aspects of the numerical steps to be addressed in later sections.

\subsubsection{Estimation of the Mean Moments of the Carrier Flow on a Mesh}~
In the present work, we have chosen to adopt a hybrid formulation in which the fluid mean fields are computed by a classical Navier-Stokes code with a turbulence model (typically a mesh-based solver using finite-volume techniques, referred to as the FV solver) and provided to the Lagrangian code (referred to as the Particle solver) to be used in the particle evolution equations, cf. Eqs.~\eqref{eq:SLM_cont}. A first advantage for doing so is that these mean fields are free from statistical noise (as they would be, should we use a stand-alone formulation where all fields are extracted from the particle set), thereby avoiding a potential source of numerical bias (interested readers are referred to detailed studies on this point~\cite{xu1999assessment,peirano2006mean}). Another interest is that, since such hybrid formulations are used to simulate disperse two-phase flows, we are resorting to a numerical formulation whose range of application encompasses also inertial particles. In that sense, the case of fluid particles considered in this work can be regarded as the limit one (when particle inertia goes to zero) of a more general situation. On the other hand, since Lagrangian PDF models for fluid particles represent a turbulence model, we are dealing with a double description of a given turbulent flow. At the numerical level, this implies that we have duplicate fields, where for example the mean velocity field is predicted by the FV solver but also by the Particle one when first-order statistics are extracted from particle velocities. To prevent different predictions for the same physical quantity, it is necessary to ensure that these duplicate fields are identical when they correspond to the same physical variable. This consistency issue constitutes an important criterion to assess the validity of the overall formulation and, more specifically, to evaluate the numerical errors related to how particle statistics are simulated.

In hybrid formulations, consistency at the discrete FV/Particle level can only be achieved if the continuous Moment/PDF description of a turbulent flow is also consistent. This point has been addressed in several works~\cite{chibbaro2011note,minier2014guidelines,minier2015lagrangian} but is worth repeating. It is also important to realize that, once particle velocities are explicitly retained in the particle state vector as is the case here, then the turbulence model corresponding to such a PDF description is a Reynolds-stress type of model~\cite{minier2014guidelines} (for more details, see a specific discussion on this issue in~\cite[section 10.3]{minier2016statistical}). Building on the well-established relations between generalized Langevin models and resulting second-order closures~\cite{pope1994relationship,pope2000turbulent,haworth_progress_2010}, this means that, if we retain the SLM in Eq.~\eqref{eq:SLM_cont_U}, a consistent hybrid formulation consists then in selecting a Rotta model in the Moment description with a constant $C_{\text{Rotta}}$ equal to $1 + 3/2\,C_0$.

\subsubsection{Interpolation of the Averaged Carrier Fields at the Position of the Particles}~

Since particles are distributed in the whole domain, they are generally not located at the positions where the mean carrier fields are estimated. This implies that an interpolation step is necessary to determine the value of these mean fields at the particle locations. Such an interpolation scheme introduces a deterministic spatial discretization error $\mathcal{E}_{\Delta x}$  which depends on the mesh and the specific interpolation scheme which is used. This numerical error has been recognized very early in particle simulations and has been the object of many studies over the years. Note that this step requires to identify (a minima) the cells in which particles are located so that there is a close connexion between the interpolation step and particle tracking. When high-order interpolation schemes are developed, they are usually based on the mean fields in the cell where each particle is contained but also the mean fields in the neighbouring cells. However, for complex geometries or unstructured meshes, estimating the position of a particle in a mesh and the relative distance to neighbouring cells is a tedious and time consuming task~\cite{lohner1990vectorized}. For this reason, the interpolation scheme used is local, i.e. based only on the mean fields associated to the cell containing the particle. It is indeed possible to estimate the mean fields at the position of the particle using a Taylor expansion around the centre of this cell. Let us write $[\Psi]_m(\vect{X})$ the $P_m$ (i.e. piece-wise polynomial) interpolation at the order $m$ of a given quantity $\Psi$ at the location of the particle $\vect{X}$ based on the value at the centre of the corresponding cell $\vect{X}^c (\vect{X})$. This $P_m$ interpolation is defined as:

\begin{equation}
    [\Psi]_m(\vect{X}) = \sum_{|\alpha| \le m}  (\vect{\partial^\alpha} \Psi)(\vect{X}^c)
    \frac{(\vect{X} - \vect{X}^c)^\alpha}{\alpha !}.
\end{equation}

In this equation the multi-index notation is used. $(\vect{X} - \vect{X}^c)^m$ is the tensor of position at the order $m$. Similarly, $\vect{\partial^m} \Psi$ is the tensor of derivation at the order $m$. It corresponds to the value at the centre of the cell at the order 0, the gradient at the first order and the Hessian matrix at the second order etc. For quantities varying in the three directions of space the number of derivative functions to compute increases quickly with the order of interpolation. For this reason, the interpolation currently used is simply a $P_0$ (i.e. piece-wise constant) interpolation based only on the value estimated at the centre of the cell. Such interpolation is coherent with the spirit of FV method where we consider the mean fields uniform within each cell. The effects and errors introduced by such an interpolation on the dynamics of the particles will be further studied in~\sref{interpolation_issue}.

\subsubsection{Temporal Integration of the Instantaneous Quantities Associated to the Particles}~

Once the mean fields at the position of the particles are estimated, the variables entering the particle state vector are updated by integrating their SDEs. In the present situation where $\vect{Z} = ( \vect{X}, \vect{U})$, this corresponds to integrating Eqs.~\eqref{eq:SLM_cont} over the selected time step $\Delta t$ with a suitable numerical scheme. This time-integration step introduces a deterministic temporal discretization error $\mathcal{E}_{\Delta t}$ which depends on the details of the integration scheme as well as on the time step $\Delta t$. The time integration scheme used in this study is the exponential scheme proposed by~\cite{minier2003weak, peirano2006mean}. Such a scheme is unconditionally stable, explicit, and exact for uniform fields. Note that the latter condition enables to avoid introducing a temporal error for the integration of a particle which remains within a cell when using a $P_0$ interpolation. When the time step is large enough, particles can cross several cells and this leads to an interplay between spatial and time discretization errors but this aspect has been treated in detail in a recent work~\cite{balvet2023time}) and this source of error is not considered here. Therefore, in the following, calculations were carried out with sufficiently small time steps to ensure that this error remains lower than the spatial numerical ones, so that the temporal error will not be discussed in this study.

\subsubsection{Estimation of the Statistics Extracted from the Set of Particles}{\label{sec:num:stat_estim}~

Once the particle-attached variables are updated, statistics can be extracted from the particle set using local Monte Carlo estimations or, in other words, locally-applied ensemble averaging. As mentioned above, in a hybrid formulation, particle statistics are not fed back into the governing SDEs and have, therefore, no direct influence on particle dynamics. They constitute, nevertheless, the observables used to assess PDF models, since we are dealing with weak approaches (whereby only statistics obtained on a representative number of particles are relevant rather than individual particle properties). To extract these statistics, we consider a volumetric partition of the computational domain and treat each small volume as an averaging bin. This means that, once a number of particles are located in a given bin, these particles are regarded as equivalent samples of the same PDF (somewhat loosely associated to the barycentre of this small volume). Then, statistics of interest are derived by typical Monte Carlo estimations (or ensemble averaging over these particles). In the more general case of inertial particles or for specific applications, different statistical weights can be attached to particles, for example their mass when simulating poly-disperse two-phase flows. In the present study, we are dealing with fluid particles representing the same fixed amount of mass and all particles have therefore the same statistical weight, by which we simply apply ensemble averaging for the Monte Carlo estimations. It is important to be aware that this amounts to making a locally homogeneity hypothesis since we assume that, in each averaging bin, we can replace the true probabilistic expectation by spatial averaging over locally-present particles
(see~\cite[sections 6.4.4 and 8.2.8]{minier2001PDF}).

This averaging process corresponds to the nearest-grid-point (NGP) method also referred as particle-in-cell (PIC), which has the advantage of remaining local and robust even in complex partitions. This formulation is coherent with the aforementioned interpolation scheme which is also local. Similarly to the interpolation issue, higher-order methods, such as the cloud-in-cell methods which are linear or piece-wise quadratic, exist but can not be easily applied for complex geometries or meshes (see~\cite{hockney_computer_1988,peirano2006mean}). Note also that, for the definition of the volumetric partition used to extract statistics from particles, it is not mandatory to use the mesh considered in the FV approach. This is however a commonly-made choice for three main practical reasons. First, it is more convenient to handle only one mesh, which makes the numerical implementation simpler since particles have to be tracked only in this mesh, thus limiting the computational cost. Second, an hypothesis made in the estimation of FV statistics of the mean fields is that the flow within each cell is statistically uniform. Then, when using $P_0$ interpolation of these fields to obtain mean field values at particle locations, it seems relevant to use the same volume elements in which these mean fields are regarded as constant (though, this point will be revisited in~\sref{interpolation_issue}). Third, it is also practical to monitor discrepancies between duplicated fields by comparing them on the same local volumes. Unless otherwise stated, only one division of the space is therefore considered in this study and the terms used to refer to the division of the domain are mesh and cells. Furthermore, the term ``ensemble average'' will refer in this study to the statistics derived from the particle set.

A few words on associated numerical errors are now in order.

Since the number of particles used to calculate statistics is finite, Monte Carlo methods introduce a zero mean statistical error $\mathcal{E}_{N}$ which depends on the number of particles in each bin. For statistically stationary cases, we can then apply a time averaging method to reduce this statistical error~\cite{xu1999assessment,muradoglu_consistent_1999}. This means that, after a given time necessary to reach a statistically stationary state, the set of particles on which the statistics are extracted is increased at each iteration by accumulating the samples. Thus after accumulating the particles during $N^{it}$ iterations, the total number of samples used for the estimation of statistics is $N^p\times N^{it}$  with $N^p$ the number of particles. This source of error is well known and will not be treated in this work. Finally a non-zero averaged spatial error $\mathcal{E}_{\widetilde{\Delta_x}}$ can appear when computing the statistics on coarse bins in which mean quantities are not uniform~\cite{viswanathan_numerical_2011}. In our case, no smoothing step is applied and the error is caused only by the spatial variation of these mean fields in a cell, as highlighted in~\sref{sp_error_stat}.

The different steps involved in a hybrid formulation are indicated in~\tref{num_meth_sum}. In this sketch, two different types of averaging operator are used. On the left part, in the moment method, the RANS averaged values are obtained on the cells making up the mesh and are noted $\overline{(.)}$. On the right part, in the PDF method, the statistics are estimated using ensemble averaging noted $\lra{(.)}$ on averaging bins. If the cells and bins are identical, the duplicated fields corresponding to the same physical quantities (mean velocity, Reynolds tensor) that are computed twice with two different operators are then provided at the same points.

\begin{figure}[h!]
    \centering
    \includegraphics[width = \linewidth]{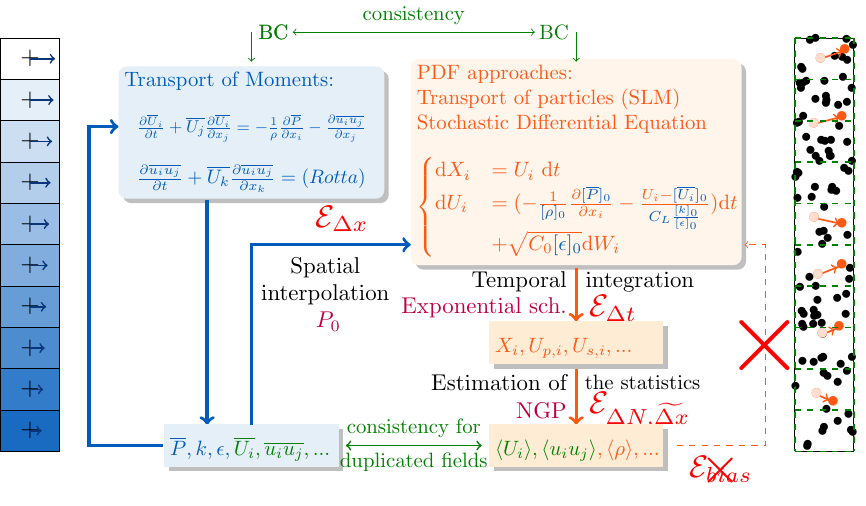}
    \caption{ Summary of different steps followed by the hybrid FV/PDF methods and numerical method considered at first (the interpolation scheme and the averaging methods are further discussed respectively in Sects.~\ref{sec:interpolation_issue} and~\ref{sec:sp_error_stat}).}
\label{tikz:num_meth_sum}
\end{figure}

\section{Verification of the Reference Wall-boundary Condition}\label{sec:verif_bc}~

The purpose of the present section is to revisit the reference wall-boundary condition (cf~\sref{wall_bc}) proposed by~\cite{dreeben_pope_1998} and~\cite{minier1999wall} to assess whether it properly represents the physics of the logarithmic zone and also to provide further numerical validation. To that effect, a turbulent 1-D infinite surface boundary-layer flow, which is characteristic of such situations, is considered. In~\sref{Surf_Boundary_layer}, we first present the analytical results obtained using the SLM model, which serves as a reference in the verification process. We then highlight that the reference wall-boundary condition enables to simulate correctly the logarithmic zone, without any modification, for both smooth and rough walls. Finally, we show that the results obtained are independent of the position of the rebound plane within the logarithmic zone, which brings in new validation results compared to previous studies.

\newpage
\subsection{Surface Boundary Layer }\label{sec:Surf_Boundary_layer}~

Leaving out thermal stratification and stability effects, the surface boundary-layer flow is the classical situation to model neutral near ground atmospheric flows. Furthermore, at high Reynolds-numbers and even in more complex situations, the flow in the vicinity of a small wall surface element can be described locally by such boundary layers. The situation considered here consists therefore in a 1-D incompressible and turbulent flow with a wall at the bottom, a constant shear stress condition at the top and periodicity in the two other directions.
The flow is driven by the shear stress imposed at the top. We impose then $\overline{u w} = - u_*^2$ where $u_*$ is the friction velocity defined as $u_* = \sqrt{\tau_{wall} / \rho}$, with $\tau_{wall}$ the shear stress at the wall.

\begin{figure}[h!]
    \centering
    \includegraphics{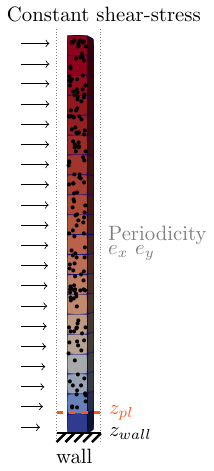}
    \caption{Scheme of the surface boundary layer studied.}
    \label{fig:domain}
\end{figure}

The high-Reynolds Navier-Stokes equation~\eref{mean_U} governing the mean velocity can be extracted from~\eref{SLM_cont_U} as:
\begin{equation}
    \frac{\lra{\dd\vect{U}}}{\dd t} = \frac{\partial \lra{\vect{U}} }{\partial t} + \mgrad(\lra{\vect{U}}) \cdot \lra{\vect{U}} + \vect{\mdiv}( \lra{\vect{u} \otimes \vect{u}}) = - \frac{1}{\rho} \mgrad \lra{P} + \vect{g}.
    \label{eq:mean_U}
\end{equation}

We consider a simple shear flow parallel to the wall, which depends only on the height: $\lra{\vect{U}} = \lra{U}(z) \vect{e_x}$ . Moreover, the diffusion terms for the second order moments are considered negligible.
 This results in a logarithmic boundary zone within which analytical results can be derived. The mean velocity profile is logarithmic and its value depends on the nature of the wall-boundary condition applied at the bottom and whose formulation for smooth or rough walls is

 \begin{subequations}
\begin{empheq}[left={\lra{U}=\empheqlbrace}]{align}
    & u_* \left( \frac{1}{\kappa} \ln(\frac{z u_*}{\nu}) + C_{log} \right)  \quad \ \text{for smooth walls ,}\label{eq:U_log_sm}
   \\
     &  \frac{u_*}{\kappa} \ln(\frac{z + z_0}{z_0})  \quad \qquad \qquad \text{for rough walls. }\label{eq:U_log_rg}
\end{empheq}
\label{eq:U_log}
 \end{subequations}

In these equations, $ \kappa$ is the Von K\'arm\'an constant equals to 0.42, $ \nu$ is the kinematic viscosity, and $C_{log}$ a constant equals to $5.2$. Two characteristic heights appear. The first one is $ \delta_\nu =  \nu /u_* $ the viscous length scale, with respect to which we define the dimensionless height $z^+ = z/ \delta_\nu = z \ u_* / \nu$ which characterizes the flow within the logarithmic boundary condition. The second one is the roughness height $z_0$ characterizing the effect of the wall roughness on the flow.

Using the SLM model and neglecting the effects of the third order terms, the equations governing the Reynolds tensor are simplified into~\eref{rij_log}:

 \begin{subequations}
\begin{alignat}{3}
  \lra{uu}: & \qquad \qquad& 2\lra{u w} \frac{\partial \lra{U}}{\partial z} = & -  \frac{2 \epsilon}{C_L}  \frac{  \langle u u  \rangle}{k} +C_0 \epsilon,  \\
  \lra{uw}: &  & \lra{w w}  \frac{\partial \lra{U}}{\partial z} = & - \frac{2 \epsilon}{C_L}  \frac{  \langle u w \rangle}{k}, \\
  \lra{vv}: &  & 0 = &   -  \frac{2 \epsilon}{C_L} \frac{  \langle v v \rangle}{k } +C_0  \epsilon,  \\
  \lra{ww}: & &  0 = &   -  \frac{2 \epsilon}{C_L} \frac{  \langle w w \rangle}{k } +C_0  \epsilon.
\end{alignat}
\label{eq:rij_log}
 \end{subequations}
Resolving these equations show that the second order moments are constant in the domain with analytical values given by~\eref{rij_log_res}:

 \begin{subequations}
        \begin{align}
              k =& \frac{1 +\frac{3}{2} C_0}{\sqrt{C_0}} u_*^2 \simeq 3.34  u_*^2\\
             \lra{u u} =&  \frac{C_0 + 2}{\sqrt{C_0}} u_*^2\simeq 2.94 u_*^2,  \\
             \lra{u w} =& -u_*^2, \\
             \lra{v v} = & \lra{w w} = \sqrt{C_0} u_*^2  \simeq 1.87  u_*^2, \\
		         \lra{u v} = & \lra{v w} = 0.
        \end{align}
        \label{eq:rij_log_res}
 \end{subequations}

\newpage
\subsection{Verification for Both Smooth and Rough Walls}\label{sec:verif_bc_sm_rg_wall}~

We now verify that the physics of the surface boundary layer is well respected using the reference wall-boundary condition presented in~\sref{wall_bc} for smooth as well as rough walls. Similar verification of the effects of the wall-boundary condition on smooth walls were proposed by~\cite{dreeben_pope_1998} and~\cite{minier1999wall} using a stand-alone approach, and by~\cite{bahlali_2020_hyb_meth} using an hybrid method. However, the case of rough walls, which is of major importance for atmospheric flows, was not considered. It is therefore interesting to extend the analysis to assess if the reference wall-boundary condition can also be applied for rough walls.

The simulations presented here were obtained with the numerical hybrid formulation introduced in~\sref{numer_methods}. In this paper, to focus on the analysis of the error introduced in the Lagrangian methods, analytic solutions described in \sref{Surf_Boundary_layer} are used for the estimation of the mean carrier fields at the centre of the cells. The computations were carried out with the open-source CFD solver code\_saturne~\cite{cs_ijfv_2004} and a uniform mesh was used with $H / \Delta z$ = 100, $H$ and  $\Delta z$ being respectively the domain and the cell heights. The natural way to shift the rebound plane away of the physical location of the wall is to extract the Lagrangian domain from the FV one. This means that the FV mesh is also used for the Lagrangian method but the boundary condition is set at an height $z_{pl}$. The particles are injected only above this plane as schematized on~\fref{domain}. We consider the flow of air in a domain of height $H$= \SI{50}{m}, with $u_* = \SI{1}{m. s^{-1}}$, thus with a Reynolds number $Re_* = H. u_*  / \nu \simeq 3.35 \ 10^{6}$. As indicated, the Lagrangian boundary condition is applied at a dimensionless height $z_{pl}^+ = z_{pl} . u_* / \nu = 1.67 . 10^5$ in the logarithmic zone. The simulations were carried out with a timestep sufficiently small compared to the lowest Lagrangian time scale seen by the particles and compared to the condition imposed by the CFL to neglect the temporal error~\cite{peirano2006mean}. Similarly, the number of particles is sufficiently high to consider that the statistical error is small compared to the spatial one~\cite{xu1999assessment,peirano2006mean}. The main source of numerical error is the spatial discretization, discussed in Sects.~\ref{sec:interpolation_issue} and~\ref{sec:sp_error_stat}.

In the following, dimensionless quantities denoted with the superscript $^*$ are plotted as a function of the dimensionless height $z/H$ (?based on the height of the domain simulated). The mean velocity and Reynolds tensor are scaled with the friction velocity $u_*$, so that $\lra{U^*} = \lra{U} /u_*$  and $\lra{u_i^*u_j^*} = \lra{u_iu_j}/u_*^2$, while the mean concentration is normalized using the concentration $\lra{C}_H$ averaged all over the domain, giving $\lra{C^*} = \lra{C} /\lra{C}_H$.

 Two main criteria are considered: first, the respect of the well-mixed criterion and, second, the respect of the theoretical values of the first and second-order velocity moments, as prescribed in~\sref{Surf_Boundary_layer}. The first condition states that, since particles represent a fixed unit of mass, their concentration must remain constant so as to ensure mass conservation and the validity of a Lagrangian stochastic approach \cite{pope_1985}. In the present case, this means that we should conserve $C^* = 1$ everywhere across the domain. As will be seen below, this criterion is always respected here since the mean pressure gradient is properly introduced in the Langevin formulation. This point has been repeatedly addressed and is now well established \cite{minier2014guidelines,bahlali_2020_hyb_meth}. We therefore concentrate mostly on discussing the second criterion.

\subsubsection{Verification on Smooth Walls}~

Results obtained using the reference wall-boundary condition and the specular one with a smooth wall are presented in~\fref{smooth_reb}.
As shown in~\fref{smooth_reb_C}, the concentration within the domain remains constant and the well-mixed criterion is respected, for both rebound conditions (the slight fluctuations are due to the inherent statistical noise of the Monte Carlo approach). This confirms that, once the mean pressure gradient is properly accounted for in the Langevin equation for particle velocities, there is no accumulation or depletion of particles even near the wall. This behaviour is not influenced by the details of the wall boundary condition applied to the particle streamwise velocity component and essentially reflects the specular reflection used for the particle vertical position. However, when we consider first- and second-order velocity moments, marked differences can be observed between the two boundary conditions. When the elastic rebound condition is used, a zero gradient condition on the velocity is actually imposed, thus not respecting the shear stress which is the driving force for the logarithmic flow. As the gradient of velocity tends towards zero, the velocity is overproduced near the rebound plane as seen in~\fref{smooth_reb_U}. Without this gradient of velocity, the production terms in~\eref{rij_log} tend to become null and this results in a tendency to have uniform turbulence near the rebound plane. Such trends are clearly apparent in~\fref{smooth_reb_rij} where the shear stress (on the left) goes to zero and the streamwise kinetic energy (on the right) tends toward the normal kinetic energy (on the middle).
On the other hand, the reference an-elastic wall-boundary condition enables to properly maintain the shear stress across the whole surface boundary layer. This is evident in~\fref{smooth_reb_rij}, where the components of the Reynolds tensor are effectively constant and in line with the values expected from~\eref{rij_log_res}. Note that we have also a proper profile of velocity in~\fref{smooth_reb_U}, i.e. linear on a semi-logarithmic scale.

\begin{figure}[htpb!]
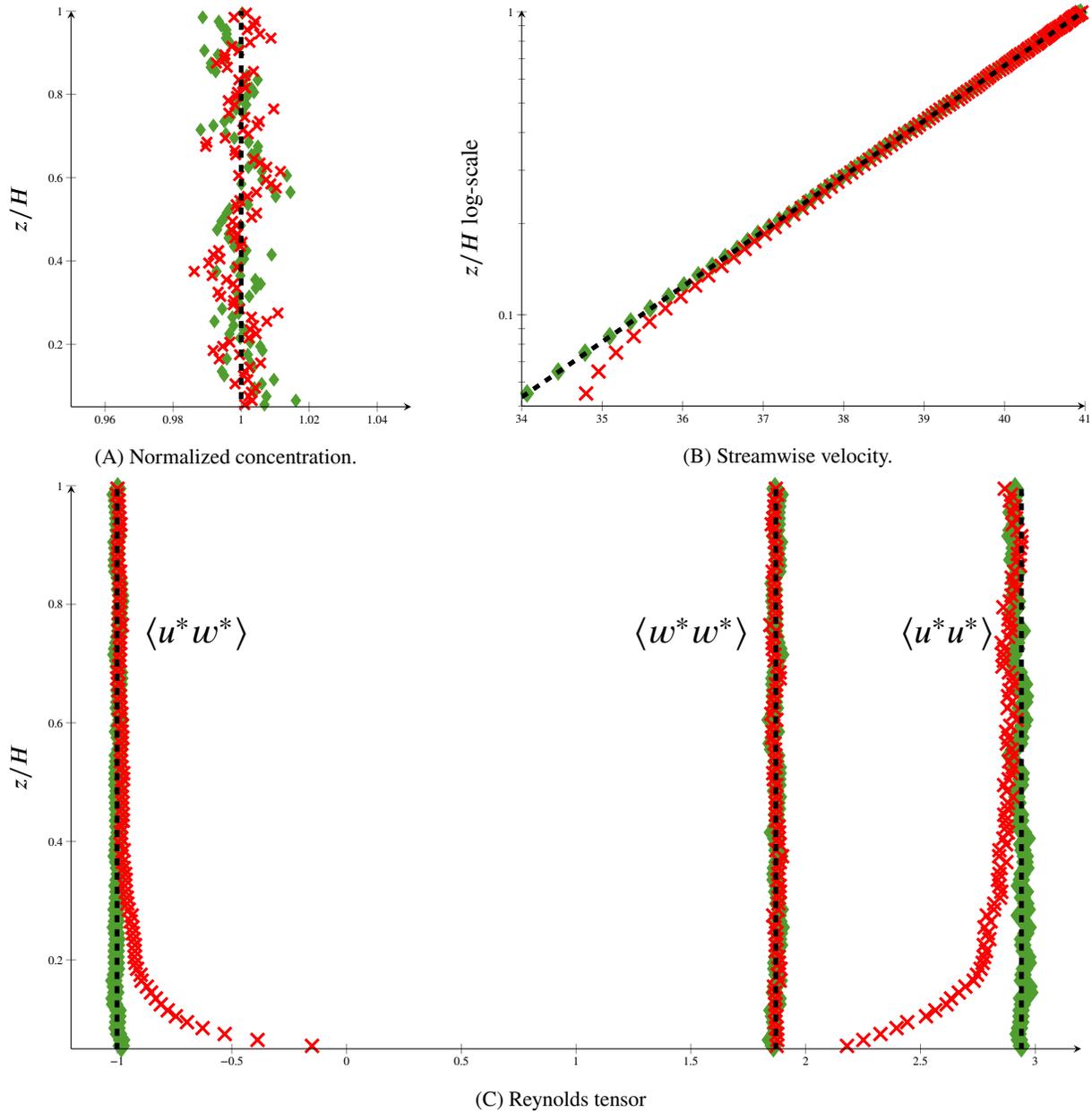

    \begin{subfigure}[]{0.4 \linewidth}
        \begin{tikzpicture}
            \input{tikz/def_init.tex}
            \def\refh{1.125}
            \def\refw{1}
            \def\isbc{1}
            \def\issmooth{1}
            \def\iscoarse{1}
            \def\isconc{1}
            \input{tikz/plot_data}
        \end{tikzpicture}
        \caption{Normalized concentration.}
        \label{fig:smooth_reb_C}
    \end{subfigure}
    \begin{subfigure}[]{0.6 \linewidth}
        \begin{tikzpicture}
            \input{tikz/def_init.tex}
            \def\refh{0.75}
            \def\refw{1}
            \def\isbc{1}
            \def\issmooth{1}
            \def\iscoarse{1}
            \def\isvel{1}
            \input{tikz/plot_data}
        \end{tikzpicture}
        \caption{Streamwise velocity.}
        \label{fig:smooth_reb_U}
    \end{subfigure}
        \begin{subfigure}[]{ \linewidth}
        \begin{tikzpicture}
            \input{tikz/def_init.tex}
            \def\refh{0.6}
            \def\refw{1}
            \def\isbc{1}
            \def\issmooth{1}
            \def\iscoarse{1}
            \def\isrij{1}
            \input{tikz/plot_data}
        \end{tikzpicture}
        \caption{Reynolds tensor}
        \label{fig:smooth_reb_rij}
    \end{subfigure}
    \caption{Vertical profiles in the surface boundary layer for smooth walls: the normalized concentration (a); the dimensionless mean streamwise velocity (b); and the four non null components of the dimensionless Reynolds tensor (c), (note that, since in the spanwise and normal direction the Reynolds tensor components are equal, only the latter one is plotted). Considering a smooth wall, two results corresponding to two distinct Lagrangian boundary conditions are compared: the reference an-elastic wall-boundary condition ({\color{greenedf}$\blacklozenge$}) and the elastic wall-boundary condition ({\color{red}$\times$}). Both respect the well-mixed criterion but only the reference wall-boundary condition enables to obtain correct mean velocity and Reynolds tensor profiles compared to the analytical solutions (black dotted lines).}
    \label{fig:smooth_reb}
\end{figure}

\subsubsection{Verification on Rough Walls}~

A similar analysis was performed with a rough wall instead of a smooth one. The roughness height considered is $z_0 = 0.1 \si{m}$. In the flow studied here, such a wall can indeed be considered as rough since $z_0^+ = z_0 u_* / \nu \sim 6.7 \ 10^3 \gg 1$. It is important to realize that the dimensionless roughness height enters only the wall-function treatment of the moment approach (thus only in the FV solver) to obtain the correct mean velocity law according to~\eref{U_log_rg}. Although this roughness height impacts the value of the mean velocity, it does not modify the shear stress which depends only on the friction velocity $u_*$ imposed on the upper face. Therefore, since the boundary condition applied on the particles, cf.~\eref{good_bc}, is built to respect a condition not directly on the mean velocity but on the shear stress, it is unchanged regardless of whether we consider smooth or rough walls. This is demonstrated by the numerical results shown in~\fref{rough_reb}, which confirm that the boundary condition used in the Lagrangian stochastic method is still valid for rough walls. Given that the shear stress is the same, these results are quite similar to the ones obtained for smooth walls and the same interpretations can be made.

\begin{figure}[htpb!]
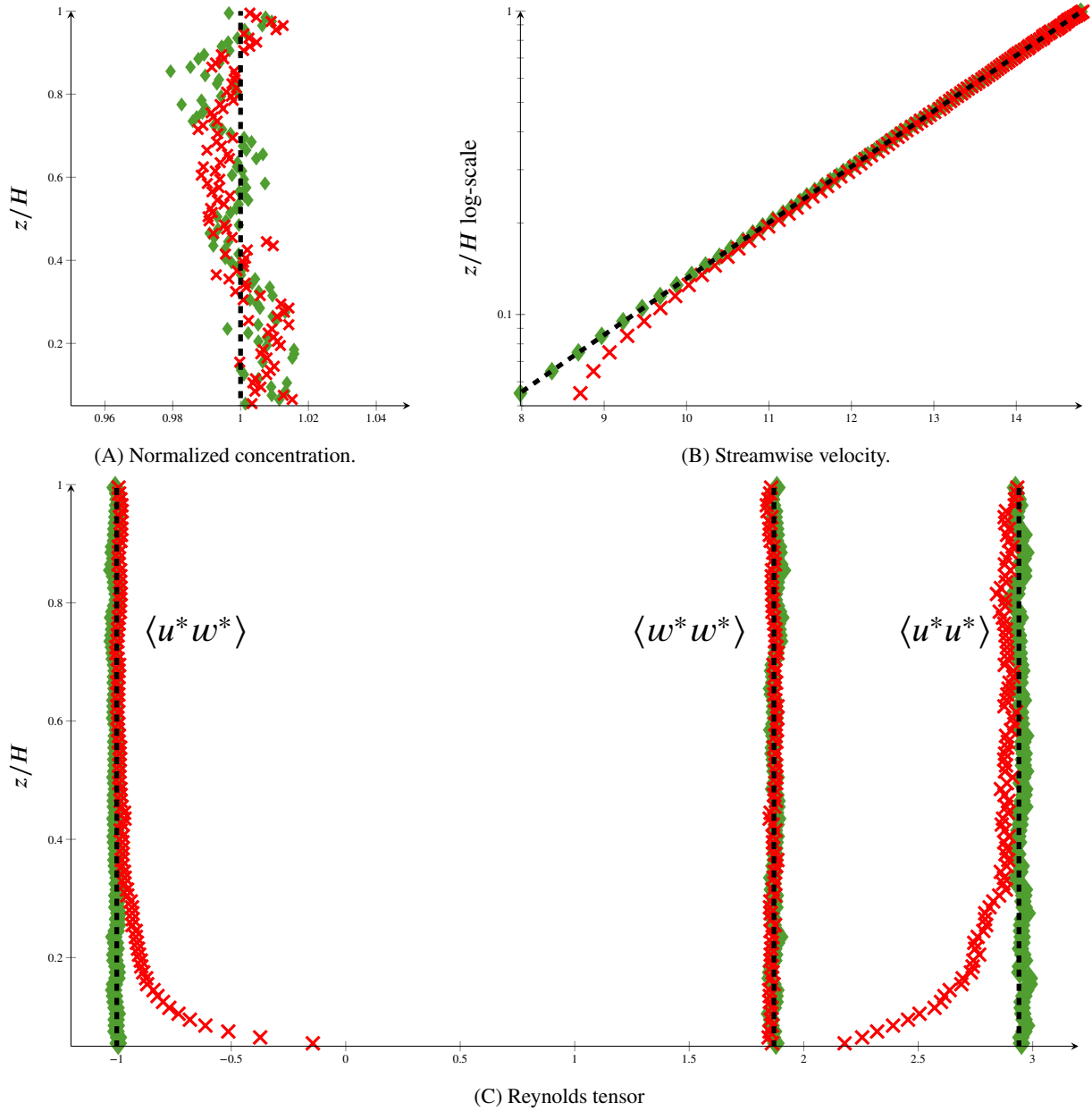

       \begin{subfigure}[]{0.4 \linewidth}
        \begin{tikzpicture}
            \input{tikz/def_init.tex}
            \def\refh{1.125}
            \def\refw{1}
            \def\isbc{1}
            \def\isrough{1}
            \def\iscoarse{1}
            \def\isconc{1}
            \input{tikz/plot_data}
        \end{tikzpicture}
        \caption{Normalized concentration.}
        \label{fig:rough_reb_C}
    \end{subfigure}
    \begin{subfigure}[]{0.6 \linewidth}
        \begin{tikzpicture}
            \input{tikz/def_init.tex}
            \def\refh{0.75}
            \def\refw{1}
            \def\isbc{1}
            \def\isrough{1}
            \def\iscoarse{1}
            \def\isvel{1}
            \input{tikz/plot_data}
        \end{tikzpicture}
        \caption{Streamwise velocity.}
        \label{fig:rough_reb_U}
    \end{subfigure}
        \begin{subfigure}[]{ \linewidth}
        \begin{tikzpicture}
            \input{tikz/def_init.tex}
            \def\refh{0.6}
            \def\isbc{1}
            \def\isrough{1}
            \def\iscoarse{1}
            \def\isrij{1}
            \input{tikz/plot_data}
        \end{tikzpicture}
        \caption{Reynolds tensor}
        \label{fig:rough_reb_rij}
    \end{subfigure}
    \caption{Vertical profiles in the surface boundary layer for rough walls: the normalized concentration (a); the dimensionless mean streamwise velocity (b); and the four non null components of the dimensionless Reynolds tensor (c) (note that in the spanwise and normal direction the Reynolds tensor components are equal, only the latter one is plotted). Considering a rough wall, two results corresponding to two distinct Lagrangian boundary conditions are compared: the reference an-elastic wall-boundary condition ({\color{greenedf}$\blacklozenge$}) and the elastic wall-boundary condition ({\color{red}$\times$}). Both respect the well-mixed criterion but only the reference wall-boundary condition enables to obtain correct mean velocity and Reynolds tensor profiles compared to the analytical solutions (black dotted lines).}
    \label{fig:rough_reb}
\end{figure}

At this stage, it is worth repeating a word of caveat: too often, a specular rebound is implemented to represent the effects of the wall~\cite{haworth_progress_2010} for high-Reynolds-number flows. As demonstrated here, this condition does not respect the physics of the logarithmic zone and, in particular, the characteristic constant shear-stress profile. For this reason it should be avoided and replaced by the reference an-elastic wall-boundary condition which, furthermore, is valid for both smooth and coarse walls. The latter rebound condition is kept for the rest of this paper.

\subsection{Independence with Respect to  \texorpdfstring{$z_{pl}^+$}{zpl+}}\label{sec:indep_zpl}~

From now on, we mainly focus on the second-order velocity moments which are the most sensible to potential sources of error.
In the previous section, the height at which the boundary condition was implemented in the logarithmic zone was chosen somewhat arbitrarily.
We now demonstrate that numerical outcomes are independent of the location of the plane at which the reference an-elastic wall-boundary condition is applied, as long as it is set in the logarithmic zone.

To this end, we consider that the rebound plane is implemented at different dimensionless heights $z_{pl}^+$. There are two ways to modify this value: either by modifying the nature of the flow through its Reynolds number, or by changing the geometrical height of the rebound plane.
The results presented in~\fref{sm_var_zpl} are based on this second method but similar results would have be obtained by modifying the Reynolds number. For these simulations only, the Reynolds number characterizing the flow is lowered to $Re_\tau = 3348$. The plane is set on different height corresponding to $z_{pl}^+ = 335$, $z_{pl}^+ = 167 $, $z_{pl}^+ = 100 $, and $z_{pl}^+ = 67$. Then, as we can see in~\fref{sm_var_zpl}, the results are independent of the choice made for the height of the rebound plane as long as it remains within the logarithmic zone.
The only difference is that by lowering the physical position of the rebound plane we can have access to information closer to the wall.

\begin{figure}[htpb!]
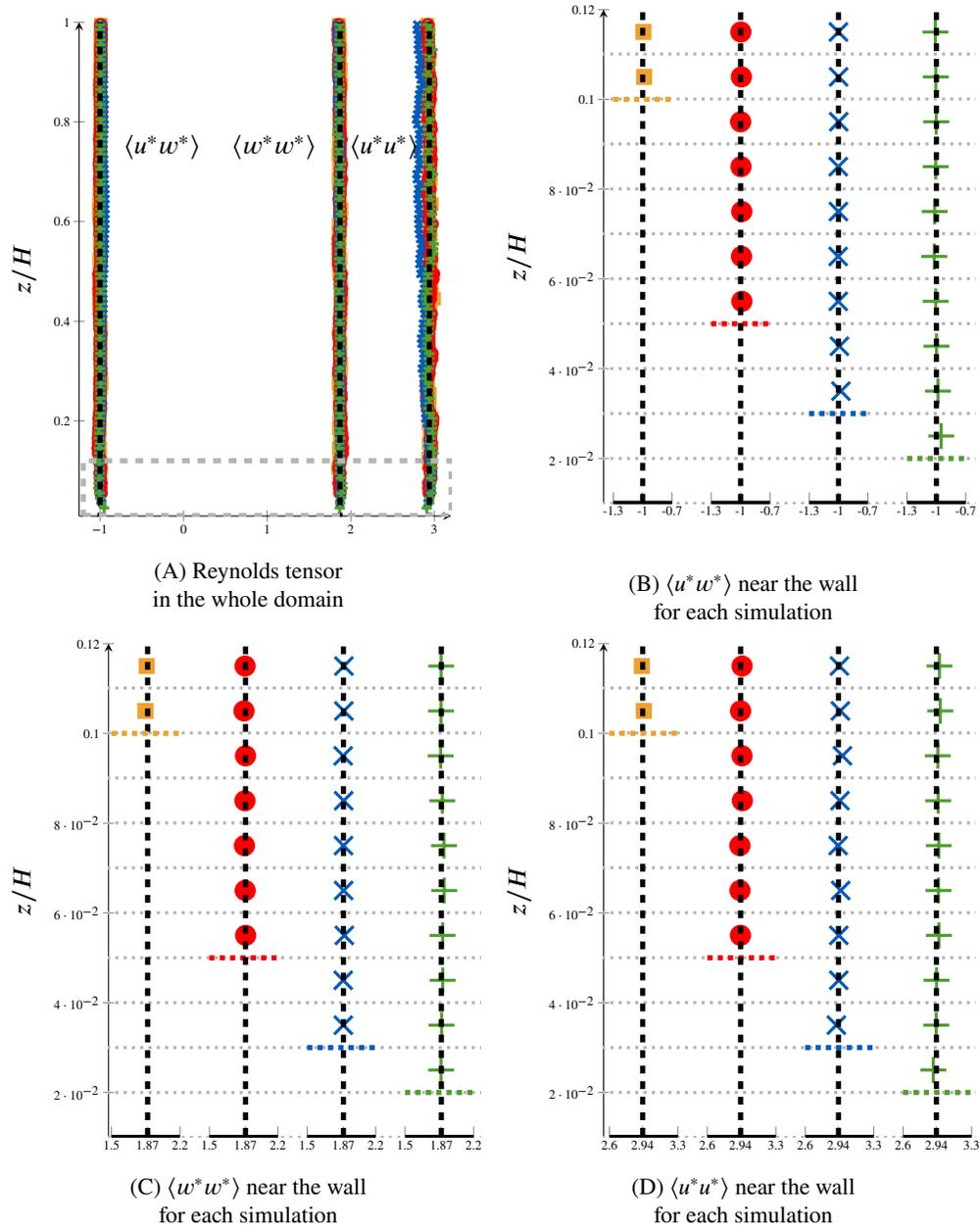

    \centering
    \begin{subfigure}[]{0.4\linewidth}
        \begin{tikzpicture}
            \input{tikz/def_init.tex}
            \def\refh{1.25}
            \def\refw{1}
            \def\issmooth{1}
            \def\iscoarse{1}
            \def\isvarzp{1}
            \def\isrij{1}
            \input{tikz/plot_data}
        \end{tikzpicture}
        \caption{ Reynolds tensor \\in the whole domain}
        \label{fig:sm_var_zpl_zoom}
    \end{subfigure}
    \begin{subfigure}[]{0.4\linewidth}
        \begin{tikzpicture}
            \input{tikz/def_init.tex}
            \def\refh{1.25}
            \def\refw{1}
            \def\issmooth{1}
            \def\iscoarse{1}
            \def\isvarzp{1}
            \def\isrij{1}
            \def\iszoom{2}
            \input{tikz/plot_data}
        \end{tikzpicture}
            \label{fig:sm_var_zpl_zoom_1}
 \caption{$\lra{u^*w^*} $ near the wall   \\ for each simulation}    \end{subfigure}
    \begin{subfigure}[]{0.4\linewidth}
        \begin{tikzpicture}
            \input{tikz/def_init.tex}
            \def\refh{1.25}
            \def\refw{1}
            \def\issmooth{1}
            \def\iscoarse{1}
            \def\isvarzp{1}
            \def\isrij{1}
            \def\iszoom{3}
            \input{tikz/plot_data}
        \end{tikzpicture}
        \caption{$\lra{w^*w^*} $ near the wall   \\ for each simulation}
    \end{subfigure}
    \begin{subfigure}[]{0.4\linewidth}
        \begin{tikzpicture}
            \input{tikz/def_init.tex}
            \def\refh{1.25}
            \def\refw{1}
            \def\issmooth{1}
            \def\iscoarse{1}
            \def\isvarzp{1}
            \def\isrij{1}
            \def\iszoom{4}
            \input{tikz/plot_data}
        \end{tikzpicture}
 \caption{$\lra{u^*u^*}  $ near the wall   \\ for each simulation}    \end{subfigure}
    \caption{
    Vertical profiles of the four non null components of the dimensionless Reynolds tensor for different boundary condition implementation heights  $z_{pl}^{+}$ (note that in the spanwise and normal direction the Reynolds tensor components are equal, only the latter one is plotted). First the vertical profiles for all components are plotted all over the domain (a). Then for each component respectively $\lra{uw}$ (b), $\lra{ww}$ (c) and $\lra{uu}$ (d), a focus is set on the few cells near the wall and for each simulation the profiles are plotted side-by-side. The implementation height considered are:  $z_{pl}^+ = 335 $ ({\color{mycolor3c}$\blacksquare$});  $z_{pl}^+ = 167 $ ({\color{mycolor3a}\Large\textbullet}); $z_{pl}^+ = 100 $ (${\color{mycolor2a}\times}$)  and $z_{pl}^+ = 67$ ({\color{mycolor1a}$+$}).
    The grey dashed box in the sub-figure (a) represents the zoomed zone near the wall. On the other sub-figures the grey dotted lines represent the FV cells and the coloured dotted lines the different boundary condition implemented. Note that the results are independent of the implementation height $z_{pl}^+$.}
    \label{fig:sm_var_zpl}
\end{figure}

To conclude, the numerical results presented in this section demonstrate the validity of the particle an-elastic wall boundary condition, \eref{good_bc}, and provide additional support compared to previous studies. Two important results complete this validation process: first, this wall boundary condition can be used, without any modification, for smooth and rough walls; and, second, the resulting profiles across the surface boundary layer are insensitive to the location of the rebound plane at which the an-elastic condition is applied provided that this rebound plane remains within the logarithmic region.
Note that wall functions in the FV formulation usually consider that the lower limit of the domain is shifted from the physical position of the wall by a distance, referred as $z_0$ for rough walls.
One can apply the PDF rebound plane directly on the boundary face of the FV mesh. This situation is considered from now on.

 Given the physical soundness of the present one-particle PDF model, we can turn our attention to the analysis of spatial numerical errors.

\newpage
\section{Interpolation of Mean Fields at Particle Positions} \label{sec:interpolation_issue}~

The purpose of this section is to introduce the issues related to the interpolation step. From now on, the fine FV volume mesh is made coarser to highlight the spatial numerical errors, we have then $H / \Delta z$ = 20. The limit of the local uniformity hypothesis is first discussed in~\sref{uniform_condition}. In~\sref{better_interpolation}, we present improved interpolation methods as well as numerical results, after selecting which mean fields are best to interpolate.

\subsection{Limitation of the Piece-wise Constant Interpolation Scheme}\label{sec:uniform_condition}~

We start by bringing out the source of error that appears when using the local uniformity hypothesis for the interpolation of mean fields at particle locations in the vicinity of the wall.

To evaluate the local uniformity hypothesis, we can compare the first order term in the Taylor decomposition to the value at the centre of the cell $\vect{X}^c$. For a variable $\Psi$, this local uniformity hypothesis requires that, everywhere in the local volume where the hypothesis is applied, we have:
\begin{equation}
        |\mgrad \Psi (\vect{X}^c)| |(X_i - X^c_i)| \ll \Psi(\vect{X}^c).
    \label{eq:cond_unif}
\end{equation}

 In our case, given that the mesh is uniform and that all the non-zero gradients increase as we get closer to the wall, the condition has just to be fulfilled at the rebound plane.
For example, this condition for the Lagrangian timescale at the rebound plane implies that:
 \begin{equation}
   \frac{2 \kappa}{ \sqrt{\mathcal{C}_0} u_* } 0.5 \Delta z  \ll \frac{2 \kappa}{ \sqrt{\mathcal{C}_0} u_* } ( 0.5 \Delta z + z_{pl}) .
   \label{eq:taylor_cond_tl}
\end{equation}

If we want to implement the rebound plane at the same location as the FV parietal law, i.e set$z_{pl} = z_{0}$, the latter condition can not be respected. Indeed, an hypothesis underlying the use of such parietal laws is to observe spatial scales larger than the shift implied by this parietal law. Therefore, we should have $2 z_{0}/$$\Delta z$$  \ll 1$.  This means that the condition \eqref{eq:taylor_cond_tl} and the uniformity hypothesis are not respected.

Furthermore, when considering uniform mean fields within a cell, the local variations of statistics and their mean gradients result also from the mixing with particles coming from nearby cells. The latter ones "keep in memory" the mean field values encountered previously during a time which is of the order of the Lagrangian time scale defined in~\eref{def_TL}. Therefore, if particles remain in a cell for a residence time larger than the local value of the Lagrangian time scale, their dynamics is mostly governed by the uniform fields within this cell, regardless of what happens in the surrounding ones and without any noticeable effects of the mean gradients. Since the mean velocity gradient is the source of the production terms in~\eref{rij_log}, under these conditions, the estimated kinetic energy of the particles would then tend toward the case of a uniform isotropic maintained turbulence (this issue is further discussed in \aref{P_0_interpol}). On the other hand, when Lagrangian time scales are small, near the interfaces between two cells, the step in the mean fields has to be bridged quickly, thus over a small distance. This yields locally to an overproduction of the estimated mean velocity gradient and of the estimated shear stress and streamwise kinetic energy. Such spurious effects are illustrated in~\tref{scheme_uniform}.

\begin{figure}[h!]
    \centering
    \includegraphics{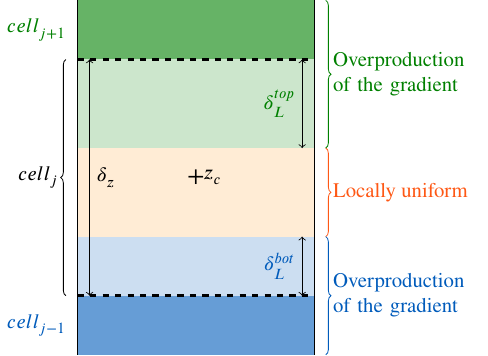}
    \caption{Illustration of the spurious evolution of the statistics within a cell when a piece-wise constant interpolation is used in a zone where the turbulent length scale $ \delta_L$  is small compared to the cell size $\Delta z$. This situation occurs near the wall. In this case the turbulent mixing is not sufficient leading to an overestimation of the gradient and production term at the faces and an underestimation of these quantities at the centre of the cells.}
    \label{tikz:scheme_uniform}
\end{figure}

Using a $P_0$ interpolation method, for this `mixing effect' to properly take place, it is thus necessary to ensure that there is no zone where particle dynamics is governed only by the uniform fields assumed in the current cell in which they are located. In order to do so, one may assert that the characteristic vertical turbulence length scale $\delta_L =\sqrt{\lra{ww}}T_L$ (the distance over which a particle "keeps memory" of carrier fields from other cells) must be large compared to the size of the cells. In this situation the evolution of $\delta_L$ is governed by:

\begin{equation}
\delta_L(z) \simeq \sqrt{\langle ww \rangle} T_L = \mathcal{C}_0^\frac{1}{4} u_*  \frac{2 \kappa}{ \sqrt{\mathcal{C}_0} u_* } z \simeq 0.61 z.
    \label{eq:memory_delta}
\end{equation}

The respect of the condition Eqs.~\eqref{eq:cond_delta_m} can give us a criteria to consider if a mesh is sufficiently fine. In the surface boundary-layer flow, it is sufficient to respect this condition at the rebound plane:

\begin{equation}
    \delta_L \simeq 0.61 (0.5 \Delta z +z_{pl}) > \Delta z.
    \label{eq:cond_delta_m}
\end{equation}

Once again, if we impose $z_{pl} = z_{0}$, this condition is not satisfied. Thus, using a $P_0$ interpolation scheme, at the centre of the cells in the immediate vicinity of the wall-boundary condition, we expect to find a zone in which particles see only the uniform properties associated to this cell. The corresponding errors are further highlighted in \aref{P_0_interpol} and solutions to limit them are discussed in~\sref{better_interpolation}.

\subsection{Improved Interpolation Methods}\label{sec:better_interpolation}~

We have seen in~\sref{uniform_condition} that, near the wall and especially if we set the rebound plane at the FV wall boundary, the $P_0$ interpolation is not valid any more and does not enable to properly recreate the particle-based mean velocity gradient and the production terms. Improved interpolation methods are thus needed. It is however useful to select first which mean mean fields are to be interpolated before addressing how this can be achieved.

\subsubsection{Selection of the Mean Fields to Interpolate}~

To select the fields to be interpolated, we can think of the mean fields terms appearing in the evolution of the moment of interest, cf.~\eref{mean_U} and~\eref{rij_log}. This corresponds to the mean velocity and its gradient, the mean pressure gradient, the Reynolds tensor and the dissipation rate. The variations of density, pressure gradient, and Reynolds tensor must be small compared to their value at the centre of the cells. This is especially true in the case of surface boundary layers where these quantities should be constant. On the other hand, near the wall, the dissipation rate and the mean velocity gradient become very large. Thus, for these fields, we can no longer estimate that only the first term in the Taylor expansion is dominant. For the mean velocity, this would result in a poor estimation of the production term and the Reynolds tensor as the ensemble-averaged velocity gradient is not handled very well near the wall with a $P_0$ interpolation. In consequence, a finer reconstruction method is needed for the mean velocity, which is to be retained as a mean field to interpolate. For similar reasons, a finer description should also be considered for the evolution of turbulent quantities. The issue is then to select which turbulent scales is best to consider. In that sense, the turbulent dissipation rate is not the best variable to interpolate due to its hyperbolic variation. Near the wall, it more convenient to consider the turbulent time scale $\tau = k/\epsilon$ which is far smoother than the turbulent dissipation rate and does not have any singular point~\cite{speziale_critical_1992}. In particular, within the logarithmic zone, $\tau$ evolves linearly with the distance to the wall whereas $\epsilon$ evolves with the inverse of this distance. Numerically, the Lagrangian time scale appears therefore as the relevant turbulent variable to interpolate. In the spirit of $k-\tau$ models, $\tau$ is then chosen to reconstruct the Lagrangian time scale and to use this quantity instead of the dissipation rate. Note that this amounts to using a $R_{ij}-T_L$ formulation instead of a $R_{ij}-\epsilon$ one. It is also worth noticing that the selection of the fields to interpolate more precisely depends on the physics considered and the variables retained in the state vector. For example, when considering the transport of a scalar $\Psi$, if we are interested in its variance or in its turbulent fluxes, the treatment of the scalar gradient in the corresponding production term requires a precise interpolation of the mean field $\lra{\Psi}$.

\newpage
\subsubsection{Interpolation Methods}~

To select interpolation methods, we first require that they respect three conditions. More precisely, a proper interpolation method should:
\begin{enumerate}[Cond. i,leftmargin=45pt]
    \item \label{item:cond_local} be local, so as to be easily implemented in complex geometries and within non homogeneous meshes (this is in line with the NGP interpolation considered);
    \item \label{item:cond_interp} provide an improved description of the mean velocity and Lagrangian time scale fields compared to the case of a uniform interpolation near the wall, so as to avoid the pitfalls presented previously;
    \item \label{item:cond_equil} respect key physical equilibrium near the wall, in particular to ensure that the production-dissipation balance is still satisfied.
\end{enumerate}

To highlight the relative importance of the two last conditions, four interpolation methods of the mean fields at the position of the particles are now introduced. They are:
\begin{enumerate}[\text{I}nterp. 1,leftmargin=45pt]
\item \label{item:interp_1}$P_0$ interpolation on the mean velocity and Lagrangian timescale fields.

    This first interpolation is the one used previously. It does not respect the conditions~\iref{cond_interp} and~\iref{cond_equil}. It should yield to the spurious effects specified in~\sref{uniform_condition} , illustrated in~\tref{scheme_uniform} and plotted in \aref{P_0_interpol} in \fref{P_0_interp_zoom}.

    \item \label{item:interp_2}$P_1$ interpolation on the mean velocity and Lagrangian timescale.

    This interpolation is the simplest and most natural one when accounting for the necessity to improve the interpolation for these two fields (see~\iref{cond_interp}). This is especially true for the Lagrangian time scale for which such an interpolation method enables to retrieve the analytical profile near the wall.
    However, within a cell, this method is not consistent with the condition~\iref{cond_equil} for the production-dissipation balance governing the dynamic of the surface boundary layer. Indeed, as the mean velocity field is linearly interpolated, its gradient is considered constant within a cell which means that the production term and the Lagrangian timescale should also be constant.

    \item \label{item:interp_3} $P_1$ interpolation on the mean velocity field and $P_0$ interpolation on the Lagrangian time scale field.

     Although less accurate than the previous one, this interpolation respects the condition~\iref{cond_equil} and the production-dissipation balance. However, it does not respect the condition~\iref{cond_interp}; as seen with~\eref{taylor_cond_tl}, the local uniformity hypothesis for the Lagrangian time scale does not hold at the wall.

    \item \label{item:interp_4} A logarithmic interpolation of the mean velocity field with a linear interpolation of the Lagrangian time scale field within wall cells and the interpolation~\iref{interp_3} everywhere else.

    This interpolation is proposed in the spirit of the wall-function treatment to retrieve both conditions~\iref{cond_interp} and~\iref{cond_equil}. Indeed, to respect the condition~\iref{cond_interp} and especially the analytical profile near the wall, a linear interpolation should be used for the Lagrangian timescale. Then, to respect the production-dissipation balance condition~\iref{cond_equil}, this choice of interpolation for the Lagrangian time scale requires to select a logarithmic interpolation for the mean velocity.
    Such an interpolation hypothesis is less justified in the whole domain and, moreover, it requires to have access to the relative position between a particle and the wall everywhere.
    Thus, to remain local and respect the condition~\iref{cond_local}, this interpolation is applied only in the cell in which the wall boundary condition is applied. In these near-wall cells, the hypothesis of a logarithmic flow is indeed consistent with the wall-function treatment used for the carrier flow.
    In the case where a cell is in contact with several wall-boundary conditions, we can consider only the interactions with the closest one to the particle. To do so, only the normal distance to the corresponding face and the shear stress associated to this face are necessary.
    Outside the immediate near-wall cells, it is necessary to consider another interpolation scheme. The one kept is the third one (i.e. \iref{interp_3}) as it respects all the three conditions outside the close vicinity of the wall.
\end{enumerate}

 In the present work, we restrict ourselves to these simple and local interpolation schemes even though more advanced and accurate interpolation methods exist~\cite{subramaniam_probability_2000,jenny_hybrid_2001,mcdermott_parabolic_2008,viswanathan_numerical_2011}.
 If one wishes to use such extended schemes, three main issues should however be carefully addressed:
 \begin{itemize}
     \item The consistency between the reconstruction of the production and dissipation terms (\iref{cond_equil}) should be assessed. Indeed, the error induced by inconsistent reconstructions can offset the gain of accuracy in the interpolation of mean fields, as shown in~\fref{interp_zoom};
     \item The implementation of such methods on 3-D unstructured meshes can turn into a daunting task;
     \item Within the scope of self-contained or stand-alone methods, the modification of the interpolation scheme is not as straightforward and can not be applied as such. Indeed, in this context, the averaging scheme used to obtained the statistics issued from the particles must be of same or higher order than the interpolation scheme used to interpolate the mean moments at the position of the particles~\cite{hockney_computer_1988}. An analysis on the averaging scheme considered would then be a prerequisite and the local NGP method used should certainly be modified.
 \end{itemize}
 For these reasons, such more complex schemes are not considered in the present work.

\subsubsection{Numerical Results}~

In order to compare the interpolations proposed above, a surface boundary layer is simulated as previously.
The simulations were carried out considering a 20 cells mesh for the finite volume calculation. A rough wall-boundary condition is implemented with a roughness height of \SI{0.1}{m}. The rebound plan is set on the same level than the FV wall-boundary condition, i.e. at a dimensionless height is $z_0^+ = 6.7\ 10^4$. Since the zone of interest is near the rebound plane, we focus on the flow obtained within the first few cells of the FV mesh. To access the spatial evolution of the extracted statistics within each FV cell in order to assess the errors induced by the different interpolation schemes, a specific treatment is applied to the statistics plotted on~\fref{interp_zoom}. The ensemble statistics are first estimated on a spatial partition of the domain 100 times finer than the FV mesh. They are then spatially averaged over ten bins, which means that in each FV cell 10 points are plotted. By doing so, we remove the spatial error introduced by the estimation of the statistics on statistically non-uniform averaging bins. This error is discussed later in~\sref{sp_error_stat}.

\begin{figure}[h!]
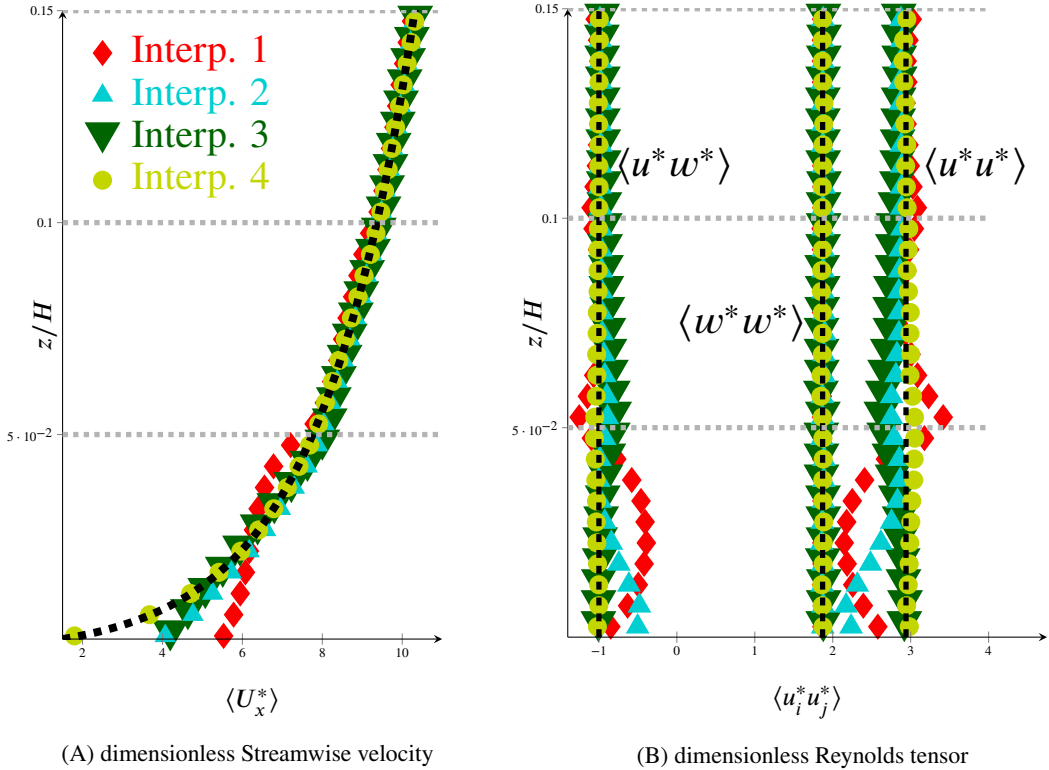

    \centering
    \begin{subfigure}[]{0.4\linewidth}
            \begin{tikzpicture}
            \input{tikz/def_init.tex}
            \def\refh{1.5}
            \def\refw{1}
            \def\isnumerr{1}
            \def\isinterp{1}
            \def\isvel{1}
            \def\isfine{2}
            \input{tikz/plot_data}
        \end{tikzpicture}
        \caption{dimensionless Streamwise velocity}
        \label{fig:interp_zoom_U}
    \end{subfigure}
    \begin{subfigure}[]{0.48\linewidth}
        \begin{tikzpicture}
            \input{tikz/def_init.tex}
            \def\refh{1.25}
            \def\refw{1}
            \def\isnumerr{1}
            \def\isinterp{1}
            \def\isrij{1}
            \def\isfine{2}
            \input{tikz/plot_data}
        \end{tikzpicture}
        \caption{dimensionless Reynolds tensor}
        \label{fig:interp_zoom_rij}
    \end{subfigure}
    \caption{
    Vertical profiles of the dimensionless mean streamwise velocity (a), the four non null components of the dimensionless Reynolds tensor (b) in the few cells near the wall (note that in the spanwise and normal direction the Reynolds tensor components are equal, only the latter one is plotted).
    The results obtained with four interpolation methods are compared:~\iref{interp_1}, which is a piece-wise constant interpolation for all mean fields({\color{red}\protect\scalebox{0.75}{$\blacklozenge$}});~\iref{interp_2}, which is a piece-wise linear interpolation for both mean velocity and mean Lagrangian time scale fields ({\color{blueledf}$\blacktriangle$});~\iref{interp_3}, which is a piece-wise linear interpolation for the mean velocity field and piece-wise constant interpolation for the mean Lagrangian time scale field ({\color{greendedf}$\blacktriangledown$});~\iref{interp_4}, which is a logarithmic interpolation for the mean velocity field and piece-wise linear interpolation for the mean Lagrangian time scale field in the cell at wall and similar to the previous interpolation otherwise ({\color{greenledf} \Large\textbullet}). These statistics are compared to the analytical solution (black dashed line). Note that a specific statistical treatment has been made to avoid the statistical error discussed on \sref{sp_error_stat}.}
    \label{fig:interp_zoom}
\end{figure}

When using the $P_0$ interpolation on all mean fields, we observe, as expected, a spurious behaviour near the wall with a uniform flow tending to appear at the centre of the first cell and overestimated fluctuations near the cell interfaces. This can be noticed by the erroneous S-shaped profile of the mean velocity within the first cell in~\fref{interp_zoom_U}. This is also quite noticeable in~\fref{interp_zoom_rij} for the estimations of the second order moments. Indeed, the streamwise components tend toward the values corresponding to an uniform maintained isotropic turbulence in the centre of the first cell while they are overpredicted at the interfaces due to overestimated gradient for the mean velocity in these areas. Those effects are well diminished by using a $P_1$ interpolation on the velocity.

However, since such $P_1$ reconstructions are in general discontinuous, we can see an error on the second order moments between the first and second cells due to the step of the mean velocity field at the cell interface. Having a more continuous profile for the Lagrangian time scale, the numerical discontinuity for the mean moments is better handled with ~\iref{interp_2} than with~\iref{interp_3}, as we can see at the interface between the two first cells. On the other hand, near the rebound plane, an underestimation of the shear stress is noticeable at the wall using~\iref{interp_2} but not using~\iref{interp_3}. \iref{interp_2} does not respect~\iref{cond_equil} for the production-dissipation balance, due to the $P_1$ interpolation of the Lagrangian time scales.
Indeed, since the velocity gradient and the production term are considered as uniform, the linear decrease of the Lagrangian time scale characterizing the turbulent structures life times is not compensated. Near the wall, this results in a spurious decrease of the shear stress and streamwise kinetic energy. Across the first cell, the consistency in the production-dissipation equilibrium, cf.~\iref{cond_equil}, seems to play a key role compared to the accuracy in the description of the Lagrangian timescale~\iref{cond_interp}. Then, \iref{interp_3} yields almost perfect predictions for the second order moment. However, near the rebound plane, there is still a noticeable discrepancy between the linear an logarithmic interpolation results. Finally, \iref{interp_4}, relying on an analytical interpolation in the first cell, seems to reproduce the exact solution for both mean velocity and Reynolds tensor. Its is thus the one proposed in the general case.

To conclude, in the vicinity of the wall a description finer than $P_0$ is necessary to properly recreate the mean velocity gradient and the production terms. It is emphasized that the consistency between the interpolation of the mean velocity and the Lagrangian time scale plays an essential role. To fulfil these conditions, the mixed interpolation method, namely~\iref{interp_4}, is proposed. In the spirit of a wall-function treatment, analytic interpolations of the mean velocity and Lagrangian time scale at particle positions are used in the cells in the immediate vicinity of the wall. Further from the wall, the simpler interpolation method, \iref{interp_3}, which consists in a piece-wise linear interpolation of mean velocity field and piece-wise constant interpolation of the Lagrangian time scale is used to respect the production-dissipation equilibrium. In order to obtain these results, a spatial average has been applied to remove spurious statistical artefacts that are studied in~\sref{sp_error_stat}.

\newpage
\section{Analysis of Statistical Bias Induced by Local Spatial Averaging}\label{sec:sp_error_stat}~

As described in~\sref{numer_methods}, either at the end of each iteration or in a post-treatment step, statistics extracted from a set of particles are estimated on a given partition of the space. For fluid particle simulations based on hybrid methods, it is worth recalling that statistics derived from the set of particles do not enter the particle evolution equations. These statistics are, however, observables used to assess particle dynamics. It is therefore of key importance to ensure that they are not biased by statistical artefacts that would result in misleading interpretations. In that sense, the main purpose of this section is to focus on the potential numerical errors appearing when estimating statistics by local averaging.

As specified in~\sref{num:stat_estim}, these errors are of two kinds. On the one hand, there is the inherent zero-average statistical noise due to Monte Carlo estimations using a finite number of particles. This well-known error converges with the square root of the number of particles and is not further discussed in the following. On the other hand, a second source of error is caused by spatial discretization when statistics are obtained over particles within small averaging bins. If the hypothesis of local homogeneity is not respected within these averaging bins, spatial errors can appear in the estimations of the covariances as demonstrated in~\sref{stat_unif}. To limit this source of error, the straightforward solution is to estimate statistics on finer bins. If the same statistics are needed on coarser bins, one may simply spatially average them in the spirit of multi-grid methods. However, this refinement process introduces multiple spatial divisions of the domain which are inconvenient for particle tracking and more time consuming. To avoid such statistical artefacts while keeping the original coarse bins, a new approach is developed in~\sref{stat_correc} which consists in correcting statistics. These new correction terms are based on an assumption about the profiles of the first order statistics within the averaging bins. Unless otherwise stated, from now on, the mean moment of the carrier flow are analytically interpolated at the position of the particles in the whole domain. This is done to focus on the error impacting the statistics without interference from issues associated to the determination of the mean moments of the carrier flow or their interpolations.

\subsection{Effect of the Non Uniformity in the Averaging Bins on the Statistics}
\label{sec:stat_unif}~

The goal in this section is to identify and quantify the source of spatial numerical error which can affect the estimation of statistics in averaging bins when the hypothesis of homogeneity, upon which probabilistic averages are replaced by spatial ones, breaks down.

\subsubsection{Effect of the Non Uniformity within the Averaging bins on the Estimator of the Moments}~

For any quantity $\Psi$, its averaged value at a position $\vect{X}$ noted $\lra{\Psi}(\vect{X})$ is approximated by the estimator of the mean noted $\lra{\Psi}_\Omega$. This estimator corresponds to the ensemble average over all the particles located within a given averaging volume $\Omega$ around the position $\vect{X}$. It converges toward the true statistic, which is the element of interest, when the averaging volume tends towards zero and the number of samples tends towards infinity, i.e. $\lra{\Psi}(\vect{X}) = \underset{\substack{\Omega \to 0 \\ n_\Omega\to \infty} }{lim}  \lra{\Psi}_\Omega $

We might wonder if $\lra{\Psi}_\Omega$ effectively represents the spatial average of the mean ($\frac{1}{\Omega}\int_\Omega \lra{\Psi}(\vect{X}) \dd \Omega$) even when $\lra{\Psi}(\vect{X})$ is not uniform in this volume.
In order to do so, let us consider a coarse averaging bin $\Omega$ containing particles uniformly distributed but in which mean moments vary (which means that the hypothesis of local homogeneity is not satisfied). Let us split this coarse volume $\Omega$ in a sufficiently high number of smaller sub-volumes $\omega$ so that statistics can be regarded as uniform within each of these sub-volumes. A scheme of such a splitting in the vertical direction is represented in \tref{compute_stat}. Note that for the sake of simplicity and clarity we consider that the mean flow varies only in one direction where the refinement is applied. Indeed, the present methodology does not depend on the number of dimension of the problem and can easily be generalized in 3-D so that it is sufficient to discuss the 1-D situation.

We assume that the number of particles $n_\omega$ within each sub-volume $\omega$ is sufficiently large to consider that the ensemble average does not suffer from statistical error. After obtaining the statistics issued from the set of particles on each of the sub-volumes, we can take the spatial average of those statistics over all the sub-volumes composing $\Omega$ as shown in \tref{compute_stat}. Under the hypothesis considered we have $\lra{\Psi}_\omega = \lra{\Psi}(\vect{X})$ and thus $\frac{1}{\Omega}\int_\Omega \lra{\Psi}(\vect{X}) \dd \Omega =\sum_{\omega \in \Omega}   \frac{n_\omega }{n_{\Omega}}\lra{\Psi}_\omega  $. For the estimator of the mean on the coarsest volume we have :

\begin{equation}
     \langle \Psi \rangle_\Omega = \frac{1}{n_\Omega} \sum_{p \in  \Omega} \Psi^p = \sum_{\omega \in \Omega}   \frac{n_\omega}{n_{\Omega}}   \underbrace{\left( \frac{1}{n_\omega}\sum_{p \in  \omega} \Psi^p\right)}_{\lra{\Psi}_\omega}
        =\frac{1}{\Omega} \int_\Omega  \lra{\Psi}(\vect{X}) \dd \Omega
\end{equation}

Thus, there is no spatial artefact on the estimation of the mean quantities even if the flow is not uniform inside a given averaging bin.

\begin{figure}[h!]
    \centering
    \includegraphics[width = \linewidth]{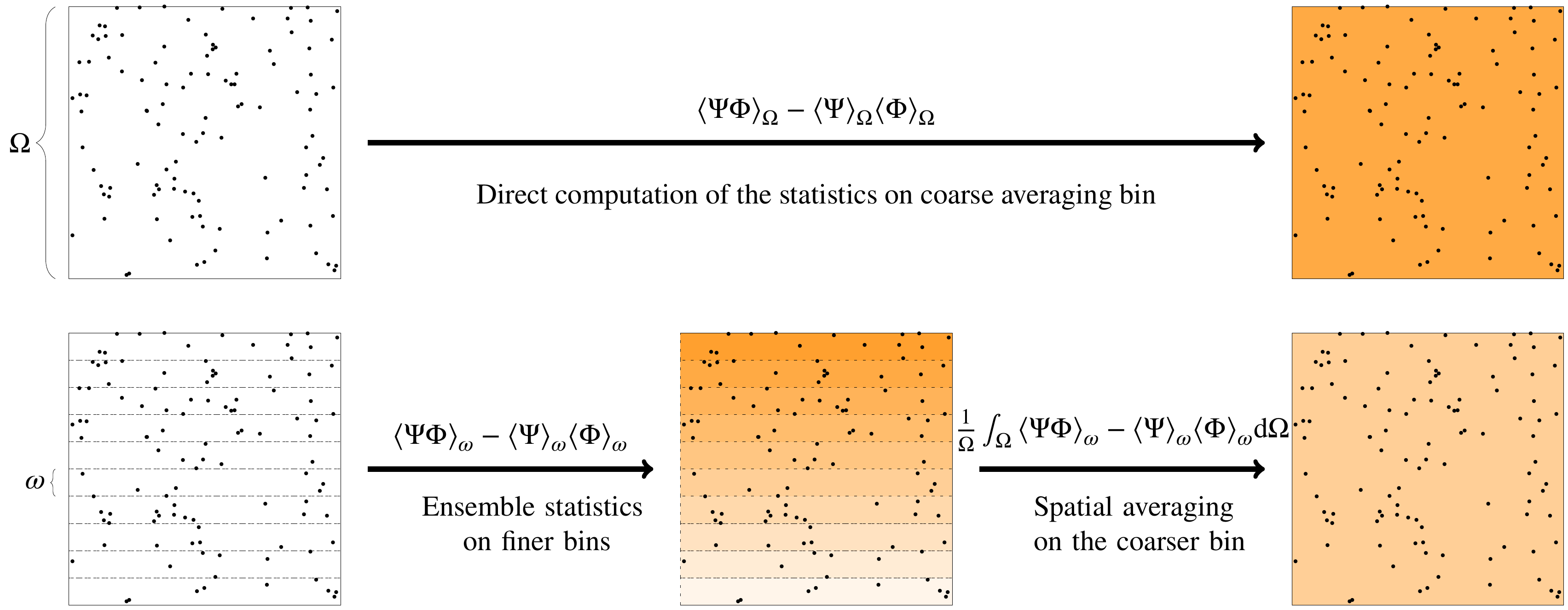}
    \caption{Illustration of the error appearing during the statistic estimation step due to the lack of uniformity of the mean field within the averaging bins. In the upper part, the covariances are estimated directly on the coarse bin $\Omega$ and are affected by numerical error. In the lower part, the refinement methods used to limit this error is depicted. First, statistics are estimated on more refined bins $\omega$ and are then spatially averaged to obtain the results on the coarser bin $\Omega$. The refinement is made only in one direction as we consider a 1-D validation case. Note that the results are affected by deterministic spatial errors only for the estimation of the covariances.}
    \label{tikz:compute_stat}
\end{figure}

We also want to check if the covariance estimator is unbiased, i.e. if the estimations of the covariance obtained on the whole volume is the same as the spatial average of the true covariance.
We have then for two quantities $\Psi$ and $ \Phi$:
\begin{align}
   \underbrace{ \lra{\Psi \Phi}_\Omega -  \lra{\Psi}_\Omega \lra{\Phi}_\Omega }_{\text{Estimator of the covariance}} &= \sum_{\omega \in \Omega}   \frac{n_\omega}{n_{\Omega}}   \underbrace{\left( \frac{1}{n_\omega}\sum_{p \in  \omega} \Psi^p \Phi^p\right)}_{\lra{\Psi \Phi}_\omega}  - \lra{\Psi}_\Omega \lra{\Phi}_\Omega \nonumber \\
    &=  \underbrace{\frac{1}{\Omega} \int_\Omega  \lra{\Psi \Phi}  - \lra{\Psi} \lra{\Phi} \dd \Omega }_{\text{True covariance of interest}}+  \underbrace{ \frac{1}{\Omega}\int_\Omega   \lra{\Psi} \lra{\Phi}  \dd\Omega- \lra{\Psi}_\Omega \lra{\Phi}_\Omega }_{\text{Bias: Spatial covariance of the mean fields}}
\label{eq:stat_cov}
\end{align}

It follows that, when the first order moments vary across the coarse averaging bin $\Omega$, the estimator of the covariances extracted directly from the whole set of particles (on the l.h.s of  \eref{stat_cov}) differs from the spatial average of the true covariances (the first term in the r.h.s term). This is emphasized by the presence of the spatial covariance of the first order moments within the coarse volume (the second r.h.s term) which is a spatial artefact caused by the non-respect of the local homogeneity hypothesis within the averaging bins.
It is clear that similar issues would occur for higher order moments. Yet, in this study, we limit ourselves to the estimation of covariance as higher order statistics are less used as observable of the flow dynamics.

 To conclude, even for correct particle dynamics, an error can impact some covariance estimators. The estimators which are biased are the ones whose corresponding first order moments vary within the averaging bins.

\subsubsection{Convergence of the Spatial Error on the Estimator of the Covariance}~

In this subsection, we estimate the magnitude of the deterministic bias introduced in \eref{stat_cov}. Since it is due to the spatial variation of mean velocity within the averaging bins, it is clear that this error depends on both the profiles of mean statistics ($\langle \Psi \rangle$, $\langle \Phi \rangle$) as well as on the size of these averaging bins. In the situation studied, as the flow is uni-axial, taking $\Psi$ and $ \Phi$ as the components of velocity, an error appears only if $\Psi = \Phi = U$. Since the spatial variance of the mean streamwise velocity is positive, if the averaging bins are too coarse, the streamwise kinetic energy is overestimated. Far from the wall, the velocity gradient is small and thus the spatial averaging bias remains low. However, as we go closer to the wall where the velocity gradient increases drastically this error becomes more important. For given particle dynamics with an analytical interpolation of the mean carrier fields at the position of the particles, this effect is demonstrated on~\fref{refin_uu}. This source of error can be preponderant, for example in this case without refinement, when the bias is of the same magnitude as the covariance itself and when the estimated covariance is twice higher than expected. It is also clear that this error converges to zero when the averaging bins are refined. Note that, since we consider a 1-D problem the refinement is made uniformly only in the direction of interest as schematized~\tref{compute_stat}.

\begin{figure}[h!]
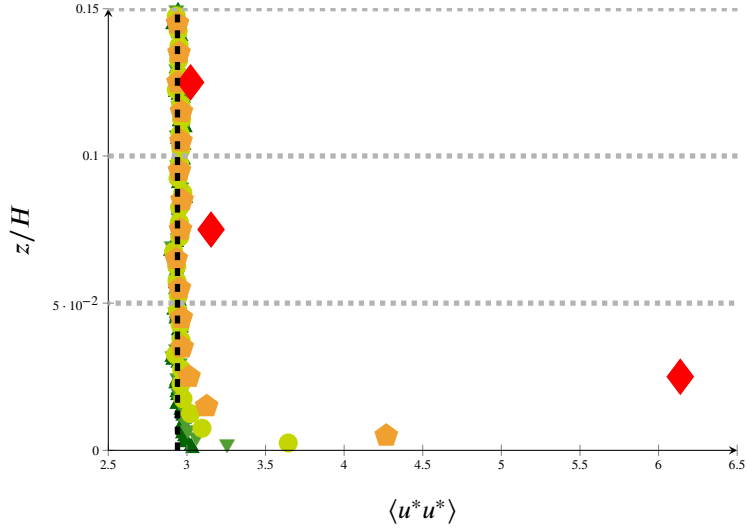

    \centering
    \begin{tikzpicture}
            \input{tikz/def_init.tex}
            \def\refh{0.45}
            \def\refw{0.6}
            \def\isnumerr{1}
            \def\isstat{1}
            \def\isrij{1}
            \input{tikz/plot_data}
        \end{tikzpicture}
    \caption{Vertical profiles of the statistical estimator of the streamwise kinetic energy in the vicinity of the wall using different spatial bins for the local averaging. Compared to the FV cells (indicated by the grey dotted lines), the bins are respectively: of the same size ({\color{red}\protect\scalebox{0.75}{$\blacklozenge$}}); 5 times finer ({\color{orangeedf}$\pentagonblack$}); 10 times finer  ({\color{greenledf} \large\textbullet}); 20 times finer ({\color{greenedf}$\blacktriangledown$}); 50 times finer ({\color{greendedf}$\blacktriangle$}). Note that these observed statistical estimators are extracted from the same particle set, i.e. they correspond to identical particle dynamics.}
    \label{fig:refin_uu}
\end{figure}

To circumvent the overestimation of covariances seen in~\sref{stat_unif}, we can first estimate statistics on a sufficiently fine partition of the domain. Once statistics are estimated on this refined partition, we can spatially average them on the coarser partition of the domain as illustrated in~\tref{compute_stat} in one dimension. With this method, the evolution of the relative error on the streamwise kinetic energy within the first coarse FV cell ($\Omega$) near the wall is plotted on~\tref{refin_conv_uu} using different refinement for the intermediate finer averaging bins ($\omega$). The coloured points correspond to the errors after averaging the profiles \fref{refin_uu} in the first cell. Two zones can be identified: a first one below a refinement factor (ratio between the cell size $\Omega$ and the bin size $\omega$) of 30 where the error converges with an order 3/2; and a second one above the refinement factor of 30 which appears as a plateau region where the other sources of numerical errors presented in~\sref{numer_methods} become preponderant (in this second zone, we can then consider the averaging bins to be small enough).

\begin{figure}[h!]
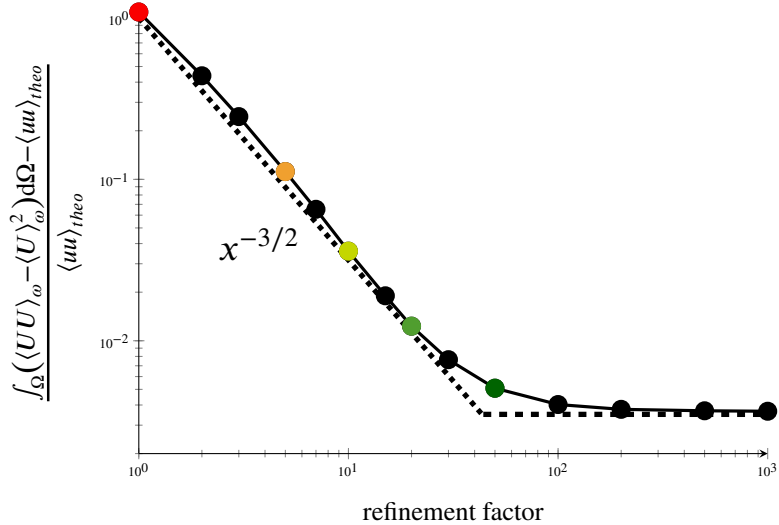

    \centering
     \begin{tikzpicture}
            \input{tikz/def_init.tex}
            \def\refh{0.45}
            \def\refw{0.6}
            \def\isnumerr{1}
            \def\isstat{2}
            \def\isrij{1}
            \input{tikz/plot_data}
        \end{tikzpicture}
    \caption{Evolution of the error on the estimator of the streamwise kinetic in the first FV cell with the refinement factor. The refinement factor corresponds to the ratio between the size of the FV cell $\Omega$ and the size of the averaging bins  $\omega$ used to estimate the intermediate statistics at first. These statistics are then spatially averaged on the FV cell.  It is observed that the error first converges with a rate $x^{-3/2}$ (dotted line) as long as the spatial error on the statistics is the main source of error (until a refinement factor around 30). Then the level of error due to the statistical noise of the Monte Carlo methods on a finite number of particles is reached and this error stagnate on a plateau (dashed lines). Each coloured dot represents the error obtained after a spatial average of the corresponding profile in \fref{refin_uu} within the first cell.
    Note that these observed statistical estimators are extracted from the same particle set, i.e. they correspond to the identical particle dynamic.}
    \label{tikz:refin_conv_uu}
\end{figure}

Such a method is applicable even for highly 3-D unstructured meshes. Indeed, each volume can always be split into sub-pyramidal volumes based on one of the faces of the original volume and on its centre of gravity. This step can be repeated to have a mesh as refined as wanted. However, tracking particles into such a finer mesh could be time consuming. Yet, in the scope of hybrid methods, the statistics issued from the set of particles do not impact their dynamics. Thus, this could be done only once as a post-processing step, so as to limit the increase of computation time. Even so, an important issue at stake is to be able to determine a criterion to ensure that the averaging bin is small enough. This criterion should provide the relative importance of this source of error compared to the other ones stated in~\sref{numer_methods}. However, criteria to quantify a priori these other sources of error do not exist in the general case. It is still possible to resort to an iterative process to determine if the averaging bins are small enough. At each iteration, the previous averaging partition of the domain is refined and the results on the new averaging partitions are compared with the results on the previous one. Once these results are similar the averaging bins can be considered small enough.

\subsection{Proposition of Correction of the Statistics on the Finite-volume Mesh Assuming Profile of the First Order Statistics within the Averaging Bins}\label{sec:stat_correc}~

To avoid iterative refinement processes and having to deal with multiple space divisions, a second idea is to directly estimate the covariance of mean moments on the coarse averaging bins $\Omega$ (e.g. the FV cells) and propose correction terms. In order to do so, we need to reconstruct the fields corresponding to the mean quantities $\lra{\Psi}$ and $\lra{\Phi}$ in the domain based on the knowledge of their values at the cell centres. These reconstructed profiles are noted $\mathcal{R}_\lra{\Psi}(\vect{X})$ and $\mathcal{R}_\lra{\Phi}(\vect{X})$, respectively, and depend on the position as well as on statistics estimated within the corresponding averaging bin $\Omega$. We can then write the correction of the covariance estimator $Cor_\Omega^{\lltriangle}\left(\mathcal{R}_\lra{\Psi}, \mathcal{R}_\lra{\Phi}\right) $ as the spatial covariance of the reconstructed mean fields:

 \begin{align*}
    Cor_\Omega^\lltriangle\left(\mathcal{R}_\lra{\Psi}, \mathcal{R}_\lra{\Phi} \right)  &= \frac{1}{\Omega}
   \int_\Omega \mathcal{R}_\lra{\Psi}(\lra{\vect{X}})\mathcal{R}_\lra{\Phi}\left(\lra{\vect{X}}\right) \dd \Omega \\
   &- \frac{1}{\Omega} \int_\Omega \mathcal{R}_\lra{\Phi}\left(\lra{\vect{X}}\right) \dd \Omega\frac{1}{\Omega} \int_\Omega \mathcal{R}_\lra{\Phi}\left(\lra{\vect{X}}\right) \dd \Omega  .
\end{align*}
This value is expected to be close to the spatial bias which is the true spatial covariance of the mean fields. Assuming that particles are uniformly distributed within the averaging bins, one may use an ensemble average over the particles located in each bin ($\lra{(.)}_\Omega$) to estimate the spatial statistics. We have then:

\begin{equation}\label{eq:gen_cor_stat}
        Cor_\Omega^\lltriangle\left(\mathcal{R}_\lra{\Psi}, \mathcal{R}_\lra{\Phi} \right)  \simeq
   \lra{ \mathcal{R}_\lra{\Psi}(\lra{\vect{X}})\mathcal{R}_\lra{\Phi}\left(\lra{\vect{X}}\right)}_\Omega - \lra{\mathcal{R}_\lra{\Phi}\left(\lra{\vect{X}}\right) }_\Omega \lra{\mathcal{R}_\lra{\Phi}\left(\lra{\vect{X}}\right)}_\Omega.
\end{equation}

At this stage, the issue is then to come up with proposals on how to reconstruct the fields corresponding to the mean moments (i.e. $\mathcal{R}_\lra{\Psi}(\vect{X})$) within the averaging bins.

\subsubsection{Proposition of Correction terms}~

In the following, three reconstruction methods are proposed. They correspond to:
\begin{enumerate}[leftmargin=50pt]

     \labitem{Reconst. unif}{item:unif_correc}
     $ \mathcal{R}^{unif}_\lra{\Psi} (\vect{X})=\lra{\Psi}_\Omega \forall \vect{X} \in \Omega$
   : no variation of the first order statistics within the averaging bins.~

    The simplest idea is to consider the mean moment constant within the averaging bin and thus the correction $Cor^{\lltriangle}_\Omega\left(\mathcal{R}_\lra{\Psi}^{unif}, \mathcal{R}_\lra{\Phi}\right)$ is null.
This is not coherent with the necessity to use finer interpolation, as discussed in~\sref{better_interpolation}. Moreover, as we can derived from~\fref{interp_zoom_U}, even a $P_0$ interpolation of the mean carrier flow does not result in a homogeneous evolution for the corresponding moment within a bin due to the mixing process discussed in~\sref{uniform_condition}.
Finer descriptions for this evolution should then be considered.~\\

\labitem{Reconst. lin}{item:lin_correc}
$ \mathcal{R}^{lin}_\lra{\Psi}$: linear variation of the first order statistics within the averaging bins.~

In the general case, one could use a Taylor expansion to estimate the evolution of the flow within a bin. The estimation of the profile for a quantity $\lra{\Psi}$ would then require to know the value of its gradient: $\mgrad_\Omega\lra{\Psi}$. We have to extract the corresponding mean gradient from the set of particles. Assuming a local uniformity of the spatial distribution of the particles in the bin, it is here proposed to estimate this quantity using the covariance of $\Psi$ and the position $\vect{X}$ and the covariance of the position as:
\begin{equation}
   \mgrad_\Omega \lra{\Psi} \simeq \Tilde{\mgrad}_\Omega \lra{\Psi} = \underbrace{ \left( \lra{\Psi \vect{X}}_\Omega -\lra{\Psi}_\Omega  \lra{\vect{X}}_\Omega \right)} _{\lra{\psi \vect{x}}_\Omega }\cdot \underbrace{\left(\lra{\vect{X} \otimes \vect{X}}_\Omega - \lra{\vect{X}}_\Omega \otimes \lra{\vect{X}}_\Omega \right)^{-1}}_{ \left(\lra{\vect{x} \otimes \vect{x}}_\Omega\right)^{-1}},
\end{equation}
with $\vect{x}$ the relative position around the cell centre, i.e. $\vect{x} = ( \vect{X} - \lra{\vect{X}})$.
The covariance of the position can be reversed since it is a symmetric definite positive tensor provided that the variance of the diagonal term is non zero. The latter condition is respected as long as there is at least 3 non coplanar particles within the averaging bin. With this estimated mean gradient $\Tilde{\mgrad}_\Omega \lra{\Psi}$, an estimated reconstruction of the mean profiles of $\lra{\Psi}$ within the bin can be built as:

\begin{equation}
      \mathcal{R}_\lra{\Psi}^{lin}(\vect{X}) = \lra{\Psi}_\Omega +\Tilde{\mgrad}_\Omega \lra{\Psi} \cdot  \vect{x}.
\end{equation}

It is then possible to estimate the correction $Cor^{\lltriangle}_\Omega\left(\mathcal{R}_\lra{\Psi}^{lin}, \mathcal{R}_\lra{\Phi}^{lin}\right)$ as:

\begin{equation}
Cor^{\lltriangle}_\Omega\left(\mathcal{R}_\lra{\Psi}^{lin}, \mathcal{R}_\lra{\Phi}^{lin}\right)= \left(\tilde{ \mgrad}_\Omega \lra{\Psi} \otimes  \tilde{\mgrad}_\Omega \lra{\Phi} \right) :   \lra{\vect{x} \otimes \vect{x}}_\Omega.
\end{equation}

\labitem{Reconst. log}{item:log_correc}
$\mathcal{R}^{log}_\lra{\Psi}$: logarithmic variation of the first order statistics within the averaging bins

Near the wall and within a surface boundary layer it is reasonable to suppose that the mean velocity follows a logarithmic profile. In order to use such a reconstruction method, we need the shear velocity $\Tilde{u}_*$. One may assume that the shear stress is constant in such zone and estimate it as:
\begin{equation}
     \Tilde{u}_* = \sqrt{|  \lra{u_\tau u_n}_\Omega|}
\end{equation}
where $u_\tau$ and $u_n$ are respectively the fluctuations of the velocity components along the streamwise and normal directions, respectively. The covariance $ \lra{u_\tau u_n}$ should not be affected by the spatial averaging error since we have near the wall $\lra{U_n} = 0 $. In the general case for a quantity $\Psi$ one may want to estimate the corresponding value $\Tilde{\psi}_*$ defined as:
\begin{equation}
         \Tilde{\psi}_* =  \frac{- \lra{\psi u_n}_\Omega}{\sqrt{|  \lra{u_\tau u_n}_\Omega|}}
\end{equation}

Based on this value, for rough walls, one can estimate $\mathcal{R}^{log}_\lra{\Psi}(\vect{X})$ as:
\begin{equation}
     \mathcal{R}^{log}_\lra{\Psi}(\vect{X}) =  \lra{\Psi}_\Omega + \frac{\Tilde{\psi}_*}{\kappa} \left(  \ln \left( d_{wall} \right) - \left\langle\ln \left(d_{wall} \right) \right\rangle_\Omega\right),
\end{equation}
where $ d_{wall}$ is the distance of the particle to the wall, typically for rough wall one may consider $ d_{wall} = X_n +z_0$.
Thus for the mean streamwise velocity, the corresponding spurious spatial covariance can be estimated as:
\begin{align*}
   & Cor^{\lltriangle}_\Omega\left(\mathcal{R}_\lra{\Psi}^{log}, \mathcal{R}_\lra{\Phi}^{log}\right)= \frac{\Tilde{\psi}_* \Tilde{\phi}_*}{\kappa^2} \Bigg( \left\langle\ln\left(d_{wall}^2\right)\right\rangle_\Omega - \left\langle\ln\left(d_{wall}\right)\right\rangle_\Omega^2\Bigg).
    \end{align*}

\end{enumerate}

Let us remark that \eref{gen_cor_stat} allows the reconstruction methods for $\lra{\Psi}$ and $\lra{\Phi}$ to differ. For example one may estimate:

\begin{equation}
    Cor^{\lltriangle}_\Omega\left(\mathcal{R}_\lra{\Psi}^{lin}, \mathcal{R}_\lra{\Phi}^{log}\right)= \frac{\Tilde{\phi}_*\Tilde{\mgrad}_\Omega\lra{\Psi}}{\kappa}   \left\langle \vect{x} \ln\left( d_{wall}\right) \right\rangle_\Omega
\end{equation}

\subsubsection{Numerical Results}~\\

To investigate the effects of the proposed correction terms in connexion with the interpolation methods, numerical results were obtained in the surface boundary-layer case. Two sets of results are presented in~\fref{corr_interp} corresponding to two slightly different particle dynamics: in~\fref{corr_interp_anal}, the mean field values at particle locations were obtained from an analytical interpolation (i.e. similar to the one used in the first cell using the~\iref{interp_4} scheme), whereas in~\fref{corr_interp_linear} a linear interpolation method (the~\iref{interp_3} scheme) was used to simulate particle dynamics. In each situation, the three different reconstruction methods for the statistical operator are then applied. It is worth stressing that, in each case, these three reconstructions correspond to the same particle dynamics.

It is seen that, in both situations, the three statistical reconstruction methods differ mostly in the first cell near the wall which, with the coarse mesh used in the simulations, is indeed where the discrepancy between the linear and logarithmic evolutions is significant (see~\fref{interp_zoom_U}). In~\fref{corr_interp_anal}, the linear assumption clearly underestimates the variation of the mean velocity within the cell and the corresponding correction term (\iref{lin_correc}) is too small. On the other hand, the correction term based on the logarithmic profile (\iref{log_correc}) enables to perfectly correct the source of error presented in~\sref{stat_unif}. In~\fref{corr_interp_linear}, the logarithmic assumption (\iref{log_correc}) strongly overestimates the variation of the mean velocity whereas the linear assumption (\iref{lin_correc}) gives now much better results. Therefore, these results show clearly that the interpolation of the mean fields at particle positions presented in~\sref{better_interpolation} and the reconstruction of these mean fields used in the corrected estimator presented in~\sref{stat_correc} must be consistent.

\begin{figure}[h!]
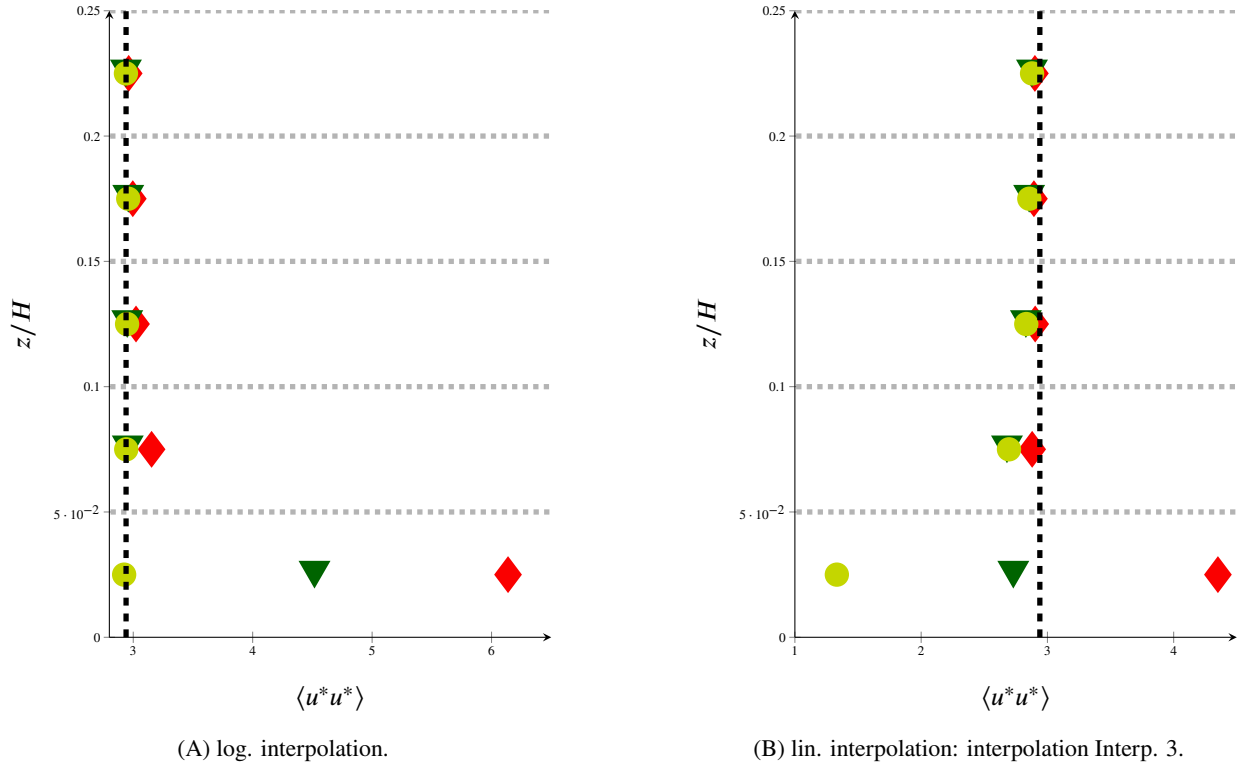

    \centering
        \begin{subfigure}[]{ 0.45\linewidth}
     \begin{tikzpicture}
            \input{tikz/def_init.tex}
            \def\refh{1.33}
            \def\refw{1}
            \def\isnumerr{1}
            \def\isstat{3}
            \def\interpid{0}
            \def\isrij{1}
            \input{tikz/plot_data}
        \end{tikzpicture}
            \caption{log. interpolation.}
            \label{fig:corr_interp_anal}
    \end{subfigure}
    \hfill
    \begin{subfigure}[]{0.45\linewidth}
         \begin{tikzpicture}
            \input{tikz/def_init.tex}
            \def\refh{1.33}
            \def\refw{1}
            \def\isnumerr{1}
            \def\isstat{3}
            \def\interpid{1}
            \def\isrij{1}
            \input{tikz/plot_data}
        \end{tikzpicture}
            \caption{lin. interpolation: interpolation~\iref{interp_3}.}
            \label{fig:corr_interp_linear}
    \end{subfigure}
    \caption{
    Vertical profiles of the statistical estimators of the streamwise kinetic energy in the vicinity of the wall for different interpolation-reconstruction combinations. The three reconstruction methods described in~\sref{stat_correc}, corresponding to three assumptions on how the ensemble mean velocity field varies in averaging bins, are compared:~\iref{unif_correc}, assuming a piece-wise constant profile ({\color{red}\protect\scalebox{0.75}{$\blacklozenge$}});~\iref{lin_correc}, assuming a piece-wise linear profile ({\color{greendedf}$\blacktriangledown$});~\iref{log_correc}, assuming a logarithmic profile ({\color{greenledf} \Large\textbullet}). The two sub-figures correspond to two different particle dynamics on which these correction terms are tested.
    In sub-figure (a), particles were simulated using the logarithmic interpolation of the mean velocity field at their positions. In that case, the linear reconstruction (\iref{lin_correc}) works better than having no correction (\iref{unif_correc}) but is not sufficient, whereas the logarithmic reconstruction (\iref{log_correc}) corrects nearly perfectly the statistical estimation.
     In sub-figure (b), particles were simulated using the piece-wise linear interpolation of the mean velocity field at their positions. In this case, the logarithmic reconstruction (\iref{log_correc}) overestimates the correction needed, whereas the linear reconstruction (\iref{lin_correc}) provides now a fairly good correction of the results obtained with no reconstruction (i.e. with~\iref{unif_correc}).}
    \label{fig:corr_interp}
\end{figure}

These conclusions are supported by additional results obtained with another interpolation method, namely \iref{interp_4} (see~\sref{better_interpolation}) which consists in a mixed interpolation scheme based on a logarithmic profile in the first cell near the wall and a linear profile otherwise. Results are presented in~\fref{corr_proposed_interp} where it is seen that nearly perfect predictions are obtained when a coherent reconstruction method is used to correct statistical estimators, i.e. using the logarithmic reconstruction in the first cell near the wall and the linear one elsewhere. Note that the requirement to have consistent methods between what appears as two adjoint operators, the interpolation method (i.e. going from the mesh to the particles) on the one hand and the one used to extract statistics (i.e. going from the particles to the mesh) on the other hand, is in line with the analysis set forth in~\cite{peirano2006mean} about similar concerns.

\begin{figure}[h!]
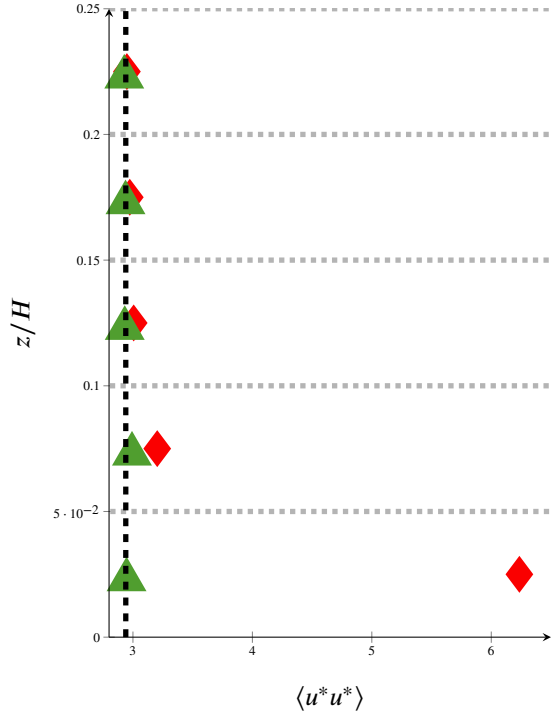

    \centering
        \begin{tikzpicture}
            \input{tikz/def_init.tex}
            \def\refh{0.6}
            \def\refw{0.45}
            \def\isnumerr{1}
            \def\isstat{3}
            \def\interpid{3}
            \def\isrij{1}
            \input{tikz/plot_data}
        \end{tikzpicture}
    \caption{Vertical profiles of the statistical estimators of the streamwise kinetic energy in the vicinity of the wall when using the method~\iref{interp_4} to interpolate the mean carrier fields at particle locations. Two results corresponding to two corrections of the statistics are compared with the analytical ones (dashed line):~\iref{unif_correc} ({\color{red}\protect\scalebox{0.75}{$\blacklozenge$}}); and, a correction coherent with the interpolation methods considered, i.e. a logarithmic reconstruction of the ensemble mean velocity $\mathcal{R}_{\lra{U}}^{log}$ in the first cell near the wall and a piece-wise linear reconstruction $\mathcal{R}_{\lra{U}}^{log}$  otherwise ({\color{greenedf}$\blacktriangle$}).}
    \label{fig:corr_proposed_interp}
\end{figure}

\newpage
\section{Conclusions and Perspectives}\label{sec:conclusions}~

In this paper, a detailed analysis to evaluate how surface boundary layers are predicted by a one-particle PDF model was carried out. This analysis was performed with two main objectives in mind.

The first main objective was to assess the validity of the wall boundary condition developed by~\cite{Dreeben1997ProbabilityDF,minier1999wall} in the spirit of the wall-function treatment of the near-wall region. Pursuing the analysis proposed by~\cite{bahlali_2020_hyb_meth}, we have shown that this reference an-elastic wall-boundary condition allows to correctly reproduce the characteristic mean-field profiles of surface boundary layers, by which we can say that the physics of such layers is indeed well captured. On the other hand, it was demonstrated that applying a specular rebound as the wall boundary condition yields non-physical results and serious discrepancies near the wall. Compared to previous studies, additional results have been obtained. Since the reference an-elastic wall-boundary condition is built so as to reproduce the shear stress within the logarithmic zone, the results put forward in~\sref{verif_bc_sm_rg_wall} prove that this condition can be applied, without any modification, for both rough and smooth walls. Another key result is that the height at which the rebound plane is implemented can be chosen arbitrarily in the logarithmic zone without impacting the statistics extracted, as demonstrated by the numerical results presented in~\sref{indep_zpl}. It is important to note that the formulation of the reference an-elastic wall boundary condition does not depend on the specific details of the Langevin model retained to simulate the velocity of fluid particles. In that sense, present results and conclusions, obtained with the simple Langevin model, are nevertheless wide range and apply to general PDF models based on particle locations and velocities.

The second main objective was to investigate spatial numerical errors in the context of hybrid FV/PDF formulations and corresponding Monte Carlo particle/mesh methods. In such hybrid simulations, it is useful to distinguish between two sources of spatial errors, depending on whether they influence particle dynamics or not. The first source of spatial error is due to the interpolation of mean fields at particle positions. These interpolated mean field values enter the Langevin equation that models the evolution of particle velocities and, therefore, affect directly particle dynamics. The results put forward in~\sref{better_interpolation} show that, to properly reproduce the local mean gradients and production terms, it is important to have a non-uniform interpolation of the mean velocity field. Furthermore, these results illustrate also a new approach introduced in the present numerical formulation, which consists in interpolating the Lagrangian time scale along with the fluid mean velocity. Numerical results indicate that the interpolation of the Lagrangian time scale should be coherent with the one used for the mean velocity so as to satisfy the production-dissipation balance at the numerical level. To fulfil these conditions, a local interpolation method has been proposed with a specific treatment in the wall cells in the spirit of the wall function. This interpolation method is based on the assumption of an established surface boundary layer within the first cell close to a wall and can be easily implemented on complex meshes. The second source of spatial error arise when particle statistics are derived by performing local Monte Carlo estimations in small volumes around a given point. In the context of hybrid FV/PDF methods, statistics for fluid particles correspond to duplicate fields and are not used in the Langevin evolution equation, which means that they do not affect particle dynamics. These statistics are mere observables but it is important to bring out potential spurious effects to avoid misleading interpretations of particle dynamics. Such statistical artefacts are linked to the breakdown of the underlying assumption of local homogeneity upon which these spatial averaging are based. Indeed, as shown in~\sref{stat_unif}, when first order statistics are not uniform within the averaging bin, spurious artefacts appear in the estimation of the corresponding covariances. To lower such effects, a first solution consists in introducing finer averaging bins. In practice, this means that we are handling two meshes, one for the fluid mean field computation in the FV solver and one to extract particle statistics in the PDF solver. In complex geometries, tracking particles in such duplicate partitions of space (or meshes) can quickly become cumbersome. To keep only one space decomposition while avoiding these potential artefacts when deriving particle statistics, new correction methods have been developed. Based on an assumed reconstruction of profiles of the first order statistics within the averaging volume, it is shown in~\sref{stat_correc} that these statistical artefacts can indeed be almost perfectly corrected. To achieve this, it is important that the interpolation scheme and the reconstruction assumed in this correction be coherent. Using this method, the statistics can then be estimated directly on the same mesh than the one used for the interpolation step.

This study has mainly focused on hybrid FV/PDF formulations. However, they can be extended to stand-alone methods, with the difference that the second source of spatial numerical error would also affect particle dynamics since we would no longer have duplicate mean fields. Furthermore, only the dynamic aspect of the flow with a very simple state vector $\vect{Z} = (\vect{X},\vect{U})$ has been taken into account in the present work. It is worth noticing that considering more complex physics, transporting additional fields such passive or active scalar, similar reasoning should be applied for the choice of the wall-boundary condition, interpolation scheme and averaging methods for the turbulent fluxes and variances of interest.

\newpage
\appendix

\section*{Acknowledgements}
G. Balvet has received a financial support by ANRT through the EDF-CIFRE contract number 2020/1387. The authors acknowledge the infrastructures at EDF R\&D and the CEREA laboratory for providing access to computational resources.

\section[Appendix1:]{Complement on the Error Induced by Piece-wise Constant Interpolation}\label{app:P_0_interpol}

The goal in the present appendix is to further discuss the influence of the interpolation methods considered on the particle dynamics. The errors occurring when using a piece-wise constant interpolation near the wall are emphasized.

Injecting the interpolated mean carrier fields at the position of the particle in the modelling of the increments of velocity \eref{SLM_cont_U},the SLM model becomes:
\begin{equation}
    \dd \vect{U}  = - \frac{1}{[\overline{\rho} ]} [\mgrad \overline{P} ]\left(t;\vect{X}(t)\right)\dd t - \frac{\vect{U} - [\vect{ \overline{U}} ](t;\vect{X}(t))}{[ \overline{T_L} ](t;\vect{X}(t))}\dd t+ \sqrt{ C_L C_0 \frac{[ \overline{k} ](t;\vect{X}(t))}{[ \overline{T_L} ](t;\vect{X}(t))}T_L(t;\vect{X}(t)) } \, \dd \vect{W}~, \label{eq:SLM_interp_U}
\end{equation}
where for any carrier fields $\Psi$, $\overline{\Psi}$ denotes the averaged value extracted from the finite volume approach and $[ \overline{\Psi} ](t;\vect{X}(t))$ its interpolation at the position of the particle $\vect{X}(t)$ at the instant  $t$.
For the sake of clarity the term $(t;\vect{X}(t))$  will be discarded from now on, yet it is important to keep in mind that generally these interpolated values are space and time dependent. The corresponding equation for the first order statistics are :

\begin{equation}
     \frac{\partial \langle U_i \rangle}{\partial t } + \langle U_k \rangle \frac{\partial \langle U_i \rangle}{\partial x_k} + \frac{\partial \langle u_k u_i \rangle}{\partial x_k}= -\frac{1}{\rho_r}\frac{\partial \overline{P}}{\partial x_i} - \frac{\lra{U_i}  -  [ \overline{U_i} ]}{[ \overline{T_L} ]}.\label{eq:interp_mean_U_i}
\end{equation}

An additional relaxation term between the mean velocity extracted from the set of particles and the interpolation of the mean carrier flow at this position appears.
For the streamwise mean velocity the equation \eref{interp_mean_U_i} becomes:

\begin{equation}
    \frac{\partial \lra{uw}}{\partial{z}} = - \frac{\lra{U}  -  [ \overline{U} ]}{[ \overline{T_L} ]}.
\end{equation}

Thus, if the mean velocity associated to the particles $(\lra{U})$ differs from the local interpolation of the mean carrier velocity fields at this position $(\overline{U})$, the local shear stress associated to the particle will not remain uniform. This effect is strengthen when approaching the wall where the Lagrangian time scale becomes small. From this equation one can obtain the equation for the particle-averaged velocity:

\begin{equation}
   \lra{U}  =  [ \overline{U} ] - \frac{\partial \lra{uw}}{\partial{z}} [ \overline{T_L} ].
\end{equation}

In a zone where the interpolated mean carrier velocity field is differentiable we can write:

\begin{equation}
   \frac{ \partial \lra{U}}{\partial z}  =  \frac{ \partial [ \overline{U} ]}{\partial z}   - \frac{\partial^2 \lra{uw}}{\partial z^2} [ \overline{T_L} ] - \frac{\partial \lra{uw}}{\partial z}  \frac{ \partial [ \overline{T_L} ]}{\partial z} .
\end{equation}

Supposing piece-wise uniform interpolation denoted $([.]_0)$, away from the faces of the cells, the gradients of the interpolated mean carrier fields are well defined and are null. We get :

\begin{equation}
    \frac{ \partial \lra{U}}{\partial z}  =   - \frac{\partial^2 \lra{uw}}{\partial z^2} [ \overline{T_L}]_0. \label{eq:grad_U_P0_interp}
\end{equation}

The derivation of the Reynolds tensor remains formally unchanged, however the shear stress being now non uniform we will also consider the turbulent diffusion terms in the equation:
\begin{equation}
  \frac{\partial \lra{u_iu_jw}}{\partial z}  = - \delta_{xj} \lra{u_i w} \frac{\partial \lra{U}}{\partial z} -  \delta_{ix} \lra{u_j w} \frac{\partial \lra{U}}{\partial z}    - \frac{2  \langle u_i u_j  \rangle}{T_L } + C_0 C_L \frac{k}{T_L}   \delta_{ij}. \label{eq:eq_uiuj_with_diff}
\end{equation}

Injecting the mean velocity gradient \eref{grad_U_P0_interp} in the shear stress equation, we have:

\begin{subequations}
\begin{align}
  \frac{\partial \langle u ww \rangle}{\partial z}  &=  \langle ww \rangle  \frac{\partial^2 \lra{uw}}{\partial z^2} [ \overline{T_L}]_0
  - 2 \frac{\langle uw \rangle  }{[ \overline{T_L}]_0},  \\
  \langle uw \rangle  =& -\frac{[ \overline{T_L}]_0}{2}  \frac{\partial \langle u ww \rangle}{\partial z} +  \langle ww \rangle  \frac{\partial^2 \lra{uw}}{\partial z^2}\frac{\left([ \overline{T_L}]_0\right)^2}{2} .\label{eq:uw_P0_interp}
\end{align}
\end{subequations}

Injecting the mean velocity gradient \eref{grad_U_P0_interp} and the shear stress \eref{uw_P0_interp} in the streamwise kinetic energy equation we get:
\begin{subequations}
\begin{align}
     \frac{\partial \langle u u w \rangle}{\partial z} & = \left(-[ \overline{T_L}]_0  \frac{\partial \langle u ww \rangle}{\partial z} +  \langle ww \rangle  \frac{\partial^2 \lra{uw}}{\partial z^2} \left([ \overline{T_L}]_0\right)^2  \right) \left( \frac{\partial^2 \lra{uw}}{\partial z^2} [ \overline{T_L}]_0\right) - 2 \frac{\langle uu \rangle  }{[ \overline{T_L}]_0} +    C_L C_0 \frac{[ \overline{k}]_0}{[ \overline{T_L}]_0},\\
     \lra{uu} &=  \underbrace{\frac{C_L C_0 [ \overline{k}]_0}{2}}_{\lra{ww}=\lra{vv} = \frac{2}{3} k^{iso}} - \frac{[ \overline{T_L}]_0}{2}  \frac{\partial \langle u u w \rangle}{\partial z} +   \frac{\left([ \overline{T_L}]_0 \right)^2 }{2} \frac{\partial^2 \lra{uw}}{\partial z^2} \left( -\frac{\partial \langle u ww \rangle}{\partial z} +  \langle ww \rangle  \frac{\partial^2 \lra{uw}}{\partial z^2} [ \overline{T_L}]_0\right). \label{eq:uu_P0_interp}
\end{align}
\end{subequations}

Assuming that we are going close to the wall where the Lagrangian time scale tends toward zero the equations \eref{grad_U_P0_interp}, \eref{uw_P0_interp}, \eref{uu_P0_interp}  imply that the particle-averaged velocity gradient and the shear stress tend towards zero whereas the streamwise kinetic energy tends toward the normal kinetic energy i.e. the one obtained for maintained isotropic turbulence. This spurious behaviour is schematized on \tref{scheme_uniform} and demonstrated on \fref{P_0_interp_zoom}.

\begin{figure}[h!]
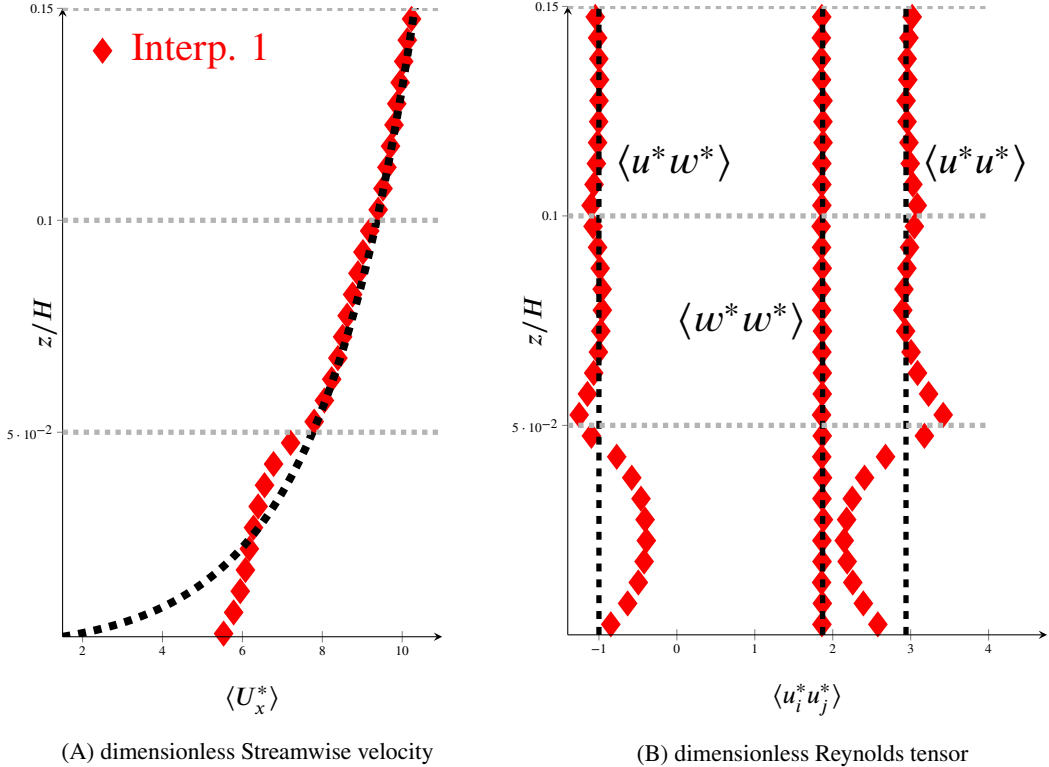

    \centering
    \begin{subfigure}[]{0.4\linewidth}
            \begin{tikzpicture}
            \input{tikz/def_init.tex}
            \def\islayer{0}
            \def\refh{1.5}
            \def\refw{1}
            \def\isnumerr{1}
            \def\isinterp{1}
            \def\isvel{1}
            \def\isfine{2}
            \input{tikz/plot_data}
        \end{tikzpicture}
        \caption{dimensionless Streamwise velocity}
        \label{fig:P0_interp_zoom_U}
    \end{subfigure}
    \begin{subfigure}[]{0.48\linewidth}
        \begin{tikzpicture}
            \input{tikz/def_init.tex}
            \def\islayer{0}
            \def\refh{1.25}
            \def\refw{1}
            \def\isnumerr{1}
            \def\isinterp{1}
            \def\isrij{1}
            \def\isfine{2}
            \input{tikz/plot_data}
        \end{tikzpicture}
        \caption{dimensionless Reynolds tensor}
        \label{fig:P_0_interp_zoom_rij}
    \end{subfigure}
    \caption{
    Vertical profiles of the dimensionless mean streamwise velocity (a), the four non null components of the dimensionless Reynolds tensor (b) in the few cells new the wall near the wall using a piece-wise uniform interpolation scheme \iref{interp_1}({\color{red}\protect\scalebox{0.75}{$\blacklozenge$}}) (note that in the spanwise and normal direction the Reynolds tensor components are equal, only the latter one is plotted). These statistics are compared with analytical solution (black dashed line). In each cell of the FV simulation (whose faces are schematized by the grey dotted lines) the statistics are first estimated into 100  finer bins. The results plotted are an agglomeration of these statistics based on a spatial average over ten bins.
    }
    \label{fig:P_0_interp_zoom}
\end{figure}

\bibliographystyle{abbrv}
\bibliography{references}
\end{document}